\documentstyle[11pt,epsfig]{article}
\begin{document}

\begin{flushright}
{hep-ph/0205xxx}\\

\end{flushright}
\vspace{1cm}
\begin{center}
{\Large \bf  Neutrinoless Double Beta Decay in Supersymmetric
Seesaw model} \vspace{.2cm}
\end{center}
\vspace{1cm}
\begin{center}
{Tai-Fu Feng$^{a,b}$\hspace{0.5cm}Xue-Qian Li$^{a,c}$
\hspace{.5cm}Yan-An Luo$^{a,c}$\\}
\vspace{.5cm}

{$^a$CCAST (World Laboratory), P.O.Box 8730,
Beijing 100080, China}\\
{$^b$Institute of High Energy Physics, Academy of Science of
China, P.O. Box
918, Beijing, 100039,  China}\\
{$^c$Department of Physics, Nankai
University, Tianjin 300070,  China}\\
\vspace{.5cm}

\end{center}
\hspace{3in}

\begin{center}
\begin{minipage}{11cm}
{\large\bf Abstract}

{\small Inspired by the recent HEIDELBERG-MOSCOW double beta decay
experiment, we discuss the neutrinoless double beta decay in the
supersymmetric seesaw model. Our numerical analysis indicates that
we can naturally explain the data of the observed neutrinoless
double beta decay, as well as that of the solar and atmospheric
neutrino experiments with at least one Majorana-like sneutrino of
middle energy scale in the model.}
\end{minipage}
\end{center}

\vspace{4mm}
\begin{center}
\end{center}


\section{Introduction}
\noindent

It is commonly considered\cite{history} that neutrinoless double
beta decay ($(\beta\beta)_{0\nu}$) is a very sensitive probe for
new physics beyond the standard model (SM). The models, such as
the Majorana mass of light neutrinos\cite{Doi}, right-handed weak
couplings involving heavy Majorana
neutrinos\cite{Halprin,Mohapatra1}, the Higgs-boson contribution
to $(\beta\beta)_{0\nu}$ in an ${\rm SU}(2)\times {\rm U}(1)$
gauge model with left-handed Majorana neutrinos\cite{ngauge}, as
well as the R-parity violation supersymmetric model that was first
proposed by Mohapatra\cite{Mohapatra2} and later studied in detail
by other groups\cite{Vergados1,Hisrch,Faessler,Uehara}, can result
in the observed neutrinoless double beta decay. Definitely, the steadily
improved experimental bounds\cite{Klapdor} on the
$(\beta\beta)_{0\nu}$ lifetime can then be translated into more
stringent limits\cite{Vergados2} for the parameter space of these
models. This would be extremely valuable information to our search
for new physics beyond SM.

Recently, a positive indication of the neutrinoless double beta
decay has been reported\cite{dbet1}. According to the announcement
of the Heidelberg group, the half-life time of
$(\beta\beta)_{0\nu}$ for nuclei $^{76}G_e$ is $(0.8 - 18.3)\times
10^{25}$ years (with the best value of $1.5\times 10^{25}$). Many
Physicists focus their attention on the work due to its important
consequences to the particle physics and astrophysics. Barger et
al. \cite{Barger} pointed out that accurate measurements of
neutrinoless double beta decay may constrain the neutrino
component of the dark matter. Combining the bound set by the CHOOZ
reactor, authors of Ref.\cite{Minakata} derived constraints on the
neutrino mixing angles; a model-independent constraint on the
neutrino mass spectrum imposed by the HEIDELBERG-MOSCOW double
beta decay experiment is also given in Ref.\cite{Xing}. Assuming a
reasonable hierarchy pattern of neutrino masses, the magnitude of
neutrinoless double beta decay is estimated in Ref.\cite{Haba};
Frigerio, Smirnov and other authors \cite{Frigerio} evaluated the
effects of various neutrino-mass-matrix structures on the
neutrinoless double beta decay. The authors of
Ref.\cite{Brahmachari,Cheng,Pakvasa,Dicus,Singh, Ahluwalia}
discussed neutrinoless double beta decay in different possible
models. In Ref.\cite{Rodejohann}, the effects of the leptonic CP
violation phase on the neutrinoless double decay have been
investigated. Meanwhile, discussions on the neutrinoless double
decay in nuclei $^{134}X_e$ are presented in Ref.\cite{Simkovic2}.

By contraries, several groups suspect if the present data can
definitely indicate a non-zero rate of $(\beta\beta)_{0\nu}$.
Aalseth et al. \cite{Aalseth} pointed out that extraction of those
signals depends on the choice of window, and some preset
conditions, such as the absence of a flat background and the
relative strength of the $^{214}B_i$ peaks etc. The authors of
Ref.\cite{Feruglio} also criticize the claim of "evidence" which
depends on the data-set choice. The most important point is that a
previous analysis\cite{062} does not find any hint towards
$(\beta\beta)_{0\nu}$ with the same data. Even so, we are inclined
to believe that the conclusion about non-zero rate of
$(\beta\beta)_{0\nu}$ is positive and then we can extract
constraint  on the parameter space of "new physics" models from
the obtained experimental data.

As it turns out\cite{236}, we cannot explain the solar neutrino,
atmospheric neutrino together with LSND collaboration observes
with three flavors of neutrinos.  In the literature, the puzzle is
addressed through the following approaches. First, one of the
three observations is simply discarded.  In the second approach  a
"sterile" neutrino\cite{058} is introduced and its existence does
not affect the decay width of $Z$ boson. In the last approach
there exists a mass-discrepancy between neutrino and anti-neutrino
which is realized by the CPT violation terms in the neutrino
sector\cite{cpt0}. Following a more common point (which is set
without any justification), we adopt the first approach while
ignoring the LSND observation due to its relatively large
experimental error (more than 3 $\sigma$) here.

If we believe the announcement of the non-zero rate of
$(\beta\beta)_{0\nu}$ and assume the light-neutrino-exchange is
the dominant mechanism for the process, by considering matrix
elements of Ref.\cite{Simkovic}, which include the contributions
from higher order terms of the nucleon current, one finds that to
fit the HEIDELBERG-MOSCOW experimental results  the effective
neutrino mass $\langle m_{\nu}\rangle=(0.11 - 0.56){\rm eV}$ (95\%
c.l.), with the best value of $0.39{\rm eV}$. Though there are a
few neutrino mixing schemes which can simultaneously explain the
data of the solar and atmospheric neutrino experiments within the
three generation model\cite{nmix}, the most favorable scenarios of
neutrino mixing lead to $\langle m_{\nu}\rangle < 0.01{\rm
eV}$\cite{nmix2}. This discrepancy implies that another mechanism
might be responsible for the neutrinoless double beta decay
process.

In this work, we will discuss the loop-induced
$(\beta\beta)_{0\nu}$ in the minimal supersymmetric extension of
the standard model with right-handed neutrinos (MSSMRN). Although
the seesaw mechanism may lead to the non-zero Majorana masses for
three light neutrinos\cite{Hirsch0}, the contributions from the
virtual Majorana neutrino-mediate diagrams are suppressed by a
small factor $\frac{\langle m_\nu\rangle}{p_F}$\cite{Bilenky},
where $\langle m_\nu\rangle$ is the "effective Majorana mass
parameter": $|\langle m_\nu\rangle|
=|U_{e1}^2m_{\nu_1}+U_{e2}^2m_{\nu_2}+U_{e3}^2m_{\nu_3}|$ and $m_
{\nu_j}$ denotes the mass of the Majorana neutrino $\nu_j$ and
$U_{ej}$ the element of the neutrino mixing matrix. $p_F\approx
100{\rm MeV}$ is the nucleon Fermi momentum. If assuming the
constraint on $<m_{\nu}>$ as $\langle m_\nu\rangle \leq 0.01{\rm
eV}$ (90\% C.L.)\cite{effmnu}, we find $\frac{\langle
m_\nu\rangle}{p_F} \leq 10^{-10}$. On the other hand, the loop
diagram contributions may play an important role in the
neutrinoless double decay if we have a relatively light
Majorana-like sneutrino. A point should be noted that a similar
computation has been performed by Hirsch {\it et al.}\cite{Hirsch}
in the supersymmetric model with non-universal soft breaking terms
and  mass terms of the Majorana-type neutrino. The model we
adopted here is a concrete realization of the general case
discussed in Ref.\cite{Hirsch}.

The paper is organized as follows. In section 2, we review the
minimal supersymmetric extension of the standard model with
right-handed neutrinos and give the notations adopted in our
analysis. In Section 3 we derive the supersymmetric contributions
to the $\Delta L=2$ effective lagrangian at the quark level. The
derivation of $\Big(\beta\beta\Big)_{0\nu}$ transition operators
and nuclear matrix elements are given in section 4. Under some
assumptions for the supersymmetric parameters, we give the numerical
analysis on $\Big(\beta\beta\Big)_{0\nu}$ decay of $^{76}G_e$ in
section 5. Our conclusions and discussions are made in section 6.
Some complicate  and tedious formulas are collected in the
appendices.

\section{The supersymmetric extension of the standard model
with right-handed neutrinos (MSSMRN) \label{nota}}

As the right-handed neutrinos are introduced into the game,
the superpotential is written as\cite{Hisano}
\begin{eqnarray}
&&{\cal W}_{_{RN}}=\mu \epsilon_{ij}\hat{H}_{i}^{1}\hat{H}_{j}^{2}+
h_{_{IJ}}^{e}\epsilon_{ij}\hat{H}_{i}^{1}\hat{L}^{I}_{j}\hat{R}^{J}
+h_{_{IJ}}^{\nu}\epsilon_{ij}\hat{H}_i^2\hat{N}^I\hat{L}_j^J
+\frac{1}{2}\hat{N}^Im_{_{IJ}}^{n}\hat{N}^J+h_{_{IJ}}^u\epsilon_{ij}
\hat{H}_i^2\hat{Q}_j^I\hat{U}^J\nonumber\\
&&\hspace{1.5cm}+h_{_{IJ}}^d\epsilon_{ij}
\hat{H}_i^1\hat{Q}_j^I\hat{D}^J\;.
\label{spot1}
\end{eqnarray}
Here, $\hat{H}^{1}$, $\hat{H}^{2}$ are the Higgs superfields,
$\hat{L}^{I}$, $\hat{Q}^{I}$ are the superfields in doublets of
the weak SU(2) group, where I=1, 2, 3 are the indices of
generations, the rest superfields $\hat{U}^{I}$, $\hat{D}^{I}$,
$\hat{N}^{I}$ and $\hat{R}^{I}$ are in singlets of the weak SU(2).
Indices i, j are contracted for the SU(2) group, and $h^{u}$,
$h^{d}$, $h^{e}$, $h^{\nu}$ are the Yukawa coupling constants. To
break supersymmetry, non-universal soft breaking terms are
introduced as
\begin{eqnarray}
&&{\cal L}_{_{RN}}^{^S}=-m_{_{H^1}}^2H_i^{1*}H_i^1
-m_{_{H^2}}^2H_i^{2*}H_i^2
-m_{_{L^{IJ}}}^2\tilde{L}_i^{I*}\tilde{L}_i^{J}
-m_{_{R^{IJ}}}^2\tilde{R}^{I*}\tilde{R}^{J}
-m_{_{N^{IJ}}}^2\tilde{N}^{I*}\tilde{N}^{J}
-m_{_{Q^{IJ}}}^2\tilde{Q}^{I*}\tilde{Q}^{J}
\nonumber \\
&&\hspace{1.3cm}
-m_{_{U^{IJ}}}^2\tilde{U}^{I*}\tilde{U}^{J}
-m_{_{D^{IJ}}}^2\tilde{D}^{I*}\tilde{D}^{J}
+(m_1\lambda_B\lambda_B+m_2\lambda_A^i\lambda_A^i
+m_3\lambda_G^a\lambda_G^a+h.c.)
+\Big[m_{_{H^{12}}}^2\epsilon_{ij}H_i^1H_j^2
\nonumber \\
&&\hspace{1.3cm}
+{\bf A}^e_{_{IJ}}\epsilon_{ij}H_{i}^{1}\tilde{L}^{I}_{j}
\tilde{R}^{J}+{\bf A}^n_{_{IJ}}\epsilon_{ij}
H_{i}^{2}\tilde{L}^{I}_{j}\tilde{N}^{J}+\frac{1}{2}\tilde{N}^{I*}
{\bf B}_{_{IJ}}^{n}\tilde{N}^{J*}+{\bf A}^d_{_{IJ}}\epsilon_{ij}
H_{i}^{1}\tilde{Q}^{I}_{j}\tilde{D}^{J}
\nonumber \\
&&\hspace{1.3cm}
+{\bf A}^u_{_{IJ}}\epsilon_{ij}
H_{i}^{2}\tilde{Q}^{I}_{j}\tilde{U}^{J}+h.c.\Big],
\label{soft}
\end{eqnarray}
where $m_{_{H^1}}^2,\;m_{_{H^2}}^2,\;m_{_{H^{12}}}^2,\;
m_{_{U^{IJ}}}^2,\;m_{_{D^{IJ}}}^2
,\;m_{_{L^{IJ}}}^2,\;m_{_{R^{IJ}}}^2$ and $m_{_{N^{IJ}}}^2$ are
parameters in unit of mass square, $m_3,\;m_2,\;m_1$ denote the
masses of $\lambda_G^a\;(a=1,\;2,\; \cdots,\;8)$,
$\lambda_A^i\;(i=1,\;2,\;3)$ and $\lambda_B$, which are the
$SU(3)\times SU(2)\times U(1)$ gauginos respectively. ${\bf B}^n$
are free parameters in unit of mass square. ${\bf
A}_{IJ}^{u},\;{\bf A}_{IJ}^{d}\;{\bf A}_{IJ}^{e},\; {\bf
A}_{IJ}^{n}\;(I,\;J=1,\;2,\;3)$ are the soft breaking parameters
that result in mass splitting between standard particles and their
supersymmetric partners. Taking into account the soft breaking
terms in Eq.(\ref{soft}), we can study the phenomenology within
the minimal supersymmetric extension of the standard model with
right-handed neutrinos(MSSMRN). The gauge symmetry $SU(2)\times
U_{_Y}(1)$ breaks down into $U_{_E}(1)$ through the nonzero vacua
of two Higgs fields. Different from the MSSM, neutrinos of three
generations obtain nonzero Majorana masses through the seesaw
mechanism \cite{Hisano}
\begin{eqnarray}
\hat{\bf m}_{\nu}=\frac{2m_{_{\rm W}}^2}{g_2^2}{\bf
U}^\dagger\left(
\begin{array}{ccc}
h^{\nu}_1& &\\&h^{\nu}_2&\\&&h^{\nu}_3\end{array}\right)({\bf
m}^n)^{-1}\left(\begin{array}{ccc} h^{\nu}_1&
&\\&h^{\nu}_2&\\&&h^{\nu}_3\end{array}\right){\bf U},
\label{neumass}
\end{eqnarray}
where the right-handed neutrino mass matrix ${\bf m}^n$ is
introduced in Eq.\ref{spot1} and the unitary matrix ${\bf U}$ is
used to diagonalize the neutrino Yukawa coupling matrix
$h^n_{_{IJ}}$.

In general, the mass matrix $\hat{\bf m}_\nu$ is still not
diagonal in the weak basis, so that we need another unitary matrix
to diagonalize $\hat{\bf m}_{\nu}$
\begin{eqnarray}
&&{\bf U_{_{M}}}^\dagger \hat{\bf m}_{\nu}{\bf U_{_{M}}}=
\frac{2m_{_{\rm W}}^2}{g_2^2}({\bf U U_{_{M}}})^\dagger\left(
\begin{array}{ccc}
h^{\nu}_1& &\\&h^{\nu}_2&\\&&h^{\nu}_3\end{array}\right)
({\bf m}^n)^{-1}\left(\begin{array}{ccc}
h^{\nu}_1& &\\&h^{\nu}_2&\\&&h^{\nu}_3\end{array}\right)
{\bf U U_{_{M}}}\nonumber \\
&&\hspace{2.2cm}=diag(m_{\nu^1},\;m_{\nu^2},\;m_{\nu^3}).
\end{eqnarray}
Thus as far as ${\bf UU_{_M}}$ is not trivially a unity matrix
{\bf I}, the mass eigenvalues are non-degenerate, one can expect
neutrino oscillation among different flavors which is a target of
current and future experiments. The mixing for $\tilde{L}_1^I$ and
$\tilde{N}^I,\;\tilde{N}^{I*}$ results in nine scalar neutrinos.
With the basis
$\Phi^T=(\tilde{L}_1^I,\;\tilde{N}^I,\;\tilde{N}^{I*})$, the
resultant mass matrix of the scalar neutrinos is written as
\begin{eqnarray}
&&\hspace{-0.8cm}\hat{\bf m}_{\tilde{\nu}}^2=\left(
\begin{array}{ccc}
\xi_{_{IJ}}&\rho_{_{IJ}}&\rho_{_{IJ}}^\dagger\\
\rho_{_{IJ}}^\dagger&
\left\{\begin{array}{l}
\frac{1}{2}(m^{n\dagger}m^n)_{IJ}+h_I^{\nu^2}\upsilon_2^2\delta_{_{IJ}}\\
+\frac{1}{2}m_{_{N^{IJ}}}^2
\end{array}\right\}
&\frac{1}{2}({\bf B}^n)_{IJ}\\
\rho_{_{IJ}}&\frac{1}{2}({\bf B}^{n\dagger})_{IJ}&
\left\{\begin{array}{l}
\frac{1}{2}(m^{n\dagger}m^n)^{IJ}+h_I^{\nu^2}\upsilon_2^2\delta_{_{IJ}}\\
+\frac{1}{2}m_{_{N^{IJ}}}^2
\end{array}\right\}
\end{array}\right),\nonumber\\
\label{snmass}
\end{eqnarray}
with
\begin{eqnarray}
&&\xi_{_{IJ}}=-{1\over 2}m_{_{\rm W}}^2\frac{C_{_\beta}^2-S_{_\beta}^2}{C_{_{\rm W}}^2}
\delta_{_{IJ}}+m_{_{L^{IJ}}}^2\;,\nonumber\\
&&\rho_{_{IJ}}=-\frac{1}{\sqrt{2}}\Big(
h_I^{\nu}\mu\upsilon_1\delta_{_{IJ}}+\upsilon_2(h^{\nu}
m^n)_{_{IJ}}-{\bf A}_{IJ}^n\upsilon_2\Big)\;, \label{xirhoij}
\end{eqnarray}
where $C_{_{\rm W}}\equiv \cos\theta_{_{\rm W}},\; S_{_{\rm
W}}\equiv \sin\theta_{_{\rm W}}$ and $\theta_{_{\rm W}}$ is the
Weinberg angle; $\upsilon_1,\;\upsilon_2$ are the nonzero vacuum
expectation values of the two  Higgs sectors and
$\tan\beta={\upsilon_2 \over \upsilon_1}$ is the ratio of the two
vacuum expectation values (hereafter we employ the symbols
$C_{_\beta}= \cos\beta,\;S_{_\beta}=\sin\beta$). The eigenvectors
$\tilde{L}_1^I,\;\tilde{N}^I,\; \tilde{N}^{I*}$ of the weak
interaction form nine sneutrino mass eigenvectors $\tilde{\nu}
^i\;(i=1,\;2,\;\cdots,\;9)$ by diagonalizing the matrix
$\hat{m}_{\tilde{\nu}}^2$:
\begin{eqnarray}
&&\tilde{\nu}^i=(Z_{\nu}^{\dagger})^{iI}\tilde{L}_1^I+
(Z_{\nu}^{\dagger})^{i(3+I)}\tilde{N}^I+
(Z_{\nu}^{\dagger})^{i(6+I)}\tilde{N}^{I*}\;,\\
\nonumber
&&Z_{\nu}^\dagger \hat{m}_{\tilde{\nu}}^2 Z_{\nu}=
diag(m_{\tilde{\nu^1}}^2,\;m_{\tilde{\nu^2}}^2,\;
\cdots,\; m_{\tilde{\nu^9}}^2),
\label{snmix}
\end{eqnarray}
where the $\tilde{L}_1^I$ are the Dirac-type sfermions and the
Lagrangian containing $\tilde{L}_1^I$  conserves the lepton
number, whereas that containing the Majorana-like scalar
$\tilde{N}^I$ violates the lepton number. As for the charged
sleptons, fields $\tilde{L}^I_2$ and $\tilde{R}^I$ mix to produce
six mass eigenvectors $\tilde{L}_i^-\;(i=1,\;2,\;\cdots\;,6)$ of
the squared-mass matrix:
\begin{eqnarray}
&&Z_L^\dagger\left(
\begin{array}{cc}
\left\{\begin{array}{l}
-{1\over 2}m_{_{\rm W}}^2\frac{(C_{_\beta}^2-S_{_\beta}^2)
(1-2C_{_{\rm W}}^2)}{C_{_{\rm W}}^2}\delta_{IJ}\\
+m_{e^I}^2\delta_{IJ}+m_{_L^{IJ}}^2\end{array}\right\} &
\frac{1}{\sqrt{2}}\Big(h_I^e\mu\upsilon_2\delta_{IJ}+
{\bf A}^e_{IJ}\upsilon_1\Big)\\
\frac{1}{\sqrt{2}}\Big(h_I^e\mu\upsilon_2\delta_{IJ}+
{\bf A}^{e\dagger}_{IJ}\upsilon_1\Big)&
\left\{\begin{array}{l}
m_{_{\rm W}}^2\frac{S_{_{\rm W}}^2(C_{_\beta}^2-S_{_\beta}^2)}
{C_{_{\rm W}}^2}\\
+m_{e^I}^2\delta_{IJ}+m_{_{L^{IJ}}}^2\end{array}\right\}
\end{array}\right)Z_L=
diag(m_{_{\tilde{L}_1^-}}^2,\;m_{_{\tilde{L}_2^-}}^2,
\cdots,\; m_{_{\tilde{L}_6^-}}^2).
\label{semix}
\end{eqnarray}

For the gaugino and scalar quark sectors, we take the notations of
Ref.\cite{notation}. Now let us turn to the phenomenology of the
supersymmetric model with right-handed neutrinos.

\section{$\Big(\beta\beta\Big)_{0\nu}$ decay in MSSMRN:
the effective Lagrangian at quark level}

In this section, considering the contributions of the model
discussed above to $\Big(\beta\beta\Big)_{0\nu}$, we derive the
relevant effective Lagrangian in terms of color-singlet currents.

\begin{figure}
\setlength{\unitlength}{1mm}
\begin{center}
\begin{picture}(230,200)(55,90)
\put(50,30){\includegraphics{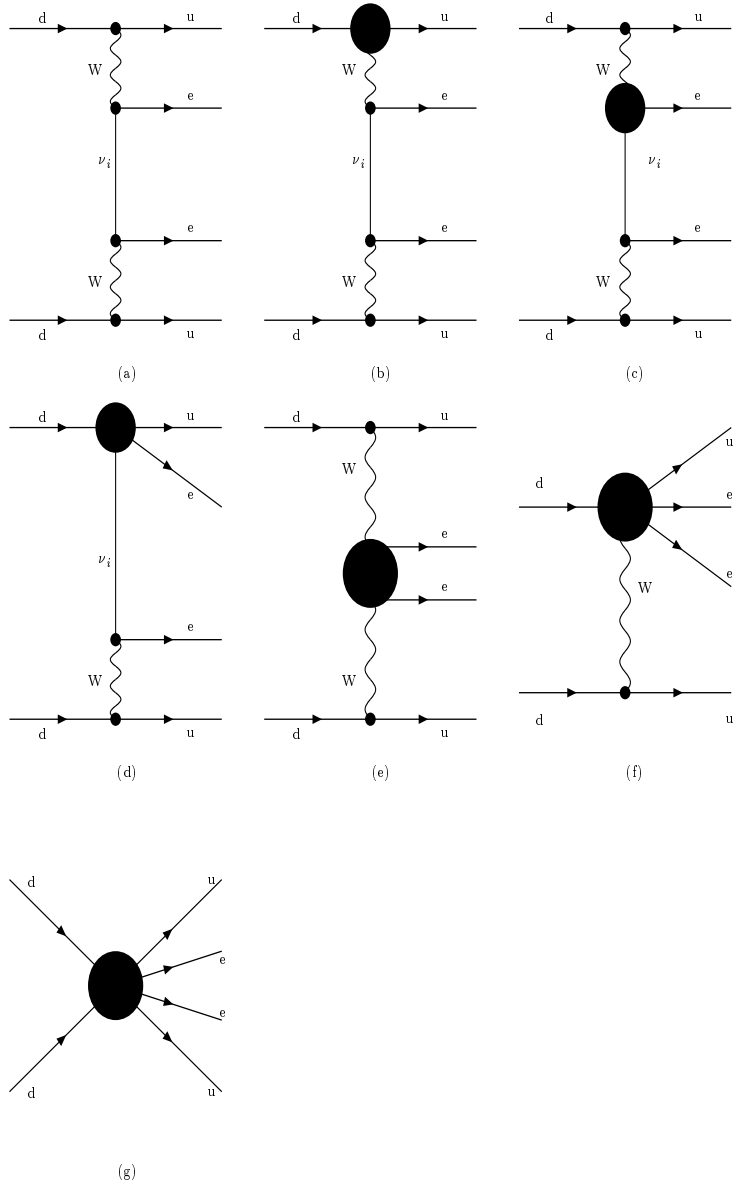}}
\end{picture}
\caption[]{The possible Feynman diagrams which contribute to
$\Big(\beta\beta\Big)_{0\nu}$ in MSSM}
\label{fig1}
\end{center}
\end{figure}

\begin{figure}
\setlength{\unitlength}{1mm}
\begin{center}
\begin{picture}(230,200)(55,90)
\put(50,30){\includegraphics{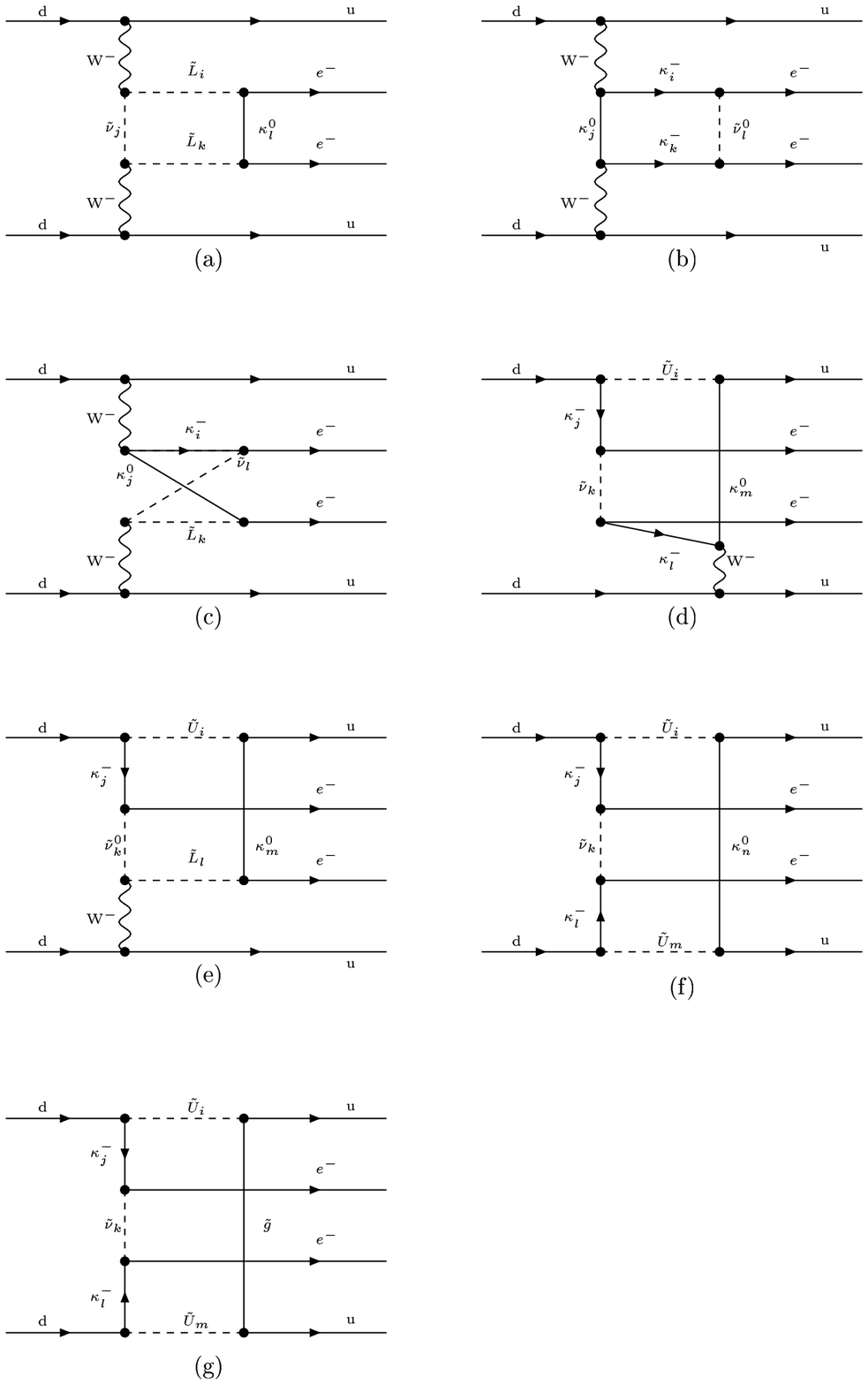}}
\end{picture}
\caption[]{The short-range contributions to
$\Big(\beta\beta\Big)_{0\nu}$ in MSSMRN (Part one)}
\label{fig2}
\end{center}
\end{figure}

\begin{figure}
\setlength{\unitlength}{1mm}
\begin{center}
\begin{picture}(230,200)(55,90)
\put(50,30){\includegraphics{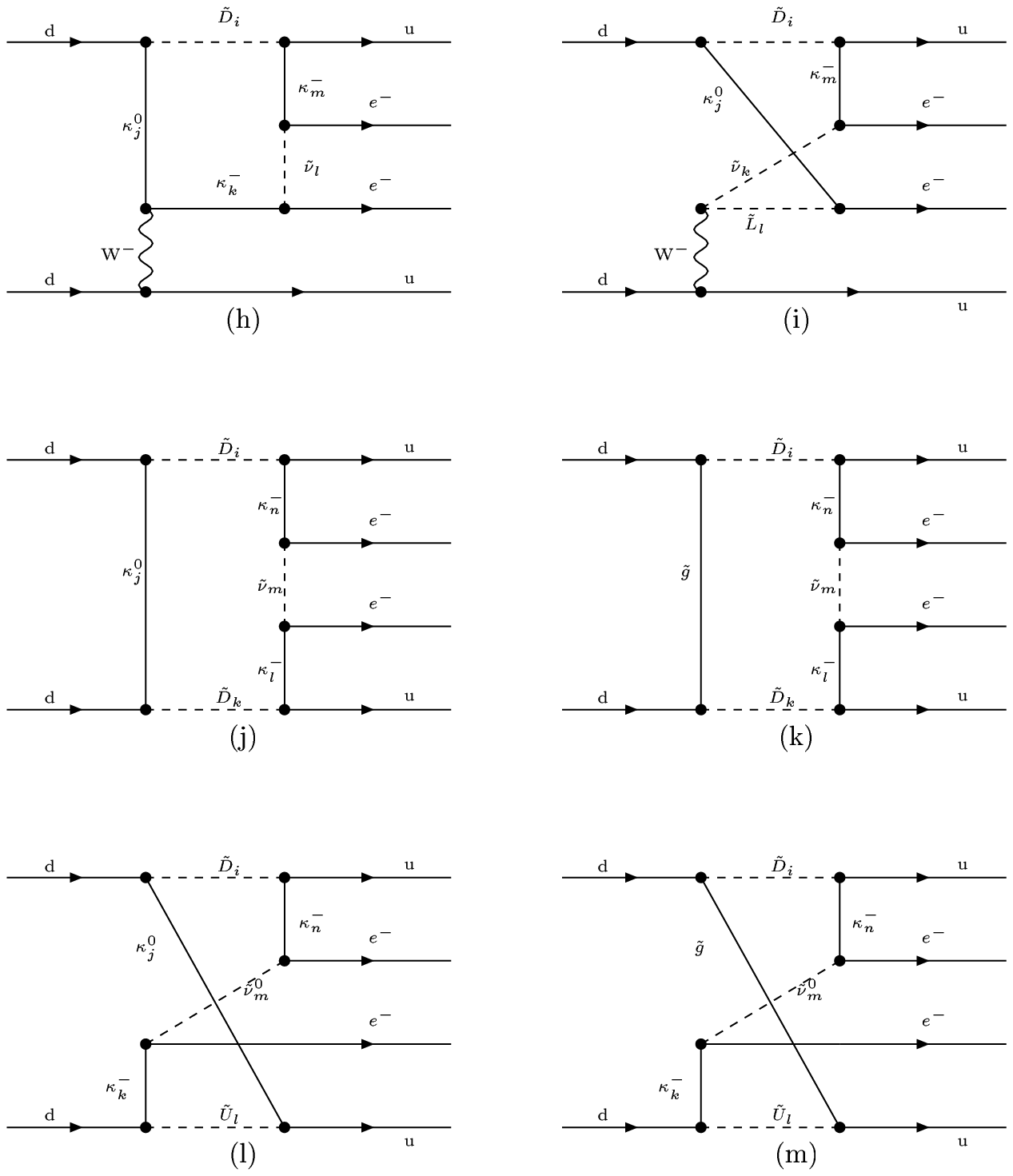}}
\end{picture}
\caption[]{The short-range contributions to
$\Big(\beta\beta\Big)_{0\nu}$ in MSSMRN (Part two)}
\label{fig3}
\end{center}
\end{figure}

Within the framework of MSSMRN, there are two sources of lepton
number violation, the contributions from the Majorana neutrinos
mediated processes and the Majorana-like sneutrinos mediated
processes to the $\Big(\beta\beta\Big)_{0\nu}$\cite{Hisrch}. At
the quark level, $\Big(\beta\beta\Big)_{0\nu}$ decay is induced by
the transition of two $d-$quarks into two $u-$quarks plus  two
electrons. All possible Feynman diagrams that induce the
$\Big(\beta\beta\Big)_{0\nu}$ decay are shown in Fig.\ref{fig1}.
As stated in the introduction, the  contributions of
Fig.\ref{fig1}(a) to $\Big(\beta\beta\Big)_{0\nu}$ decay is
suppressed by a factor $\frac{\langle m_\nu\rangle}{p_F}$.
Besides this suppression factor, the contributions of
Fig.\ref{fig1}(b,c,d) to $\Big(\beta\beta\Big)_{0\nu}$  still
suffer from the typical loop integration suppression. In contrast
to the SM case, the supersymmetric contributions are different.
Three classes of Feynman diagrams are given in
Fig.\ref{fig1}(e,f,g). It is noted that the  supersymmetric
contributions only receive the loop integration suppression which
indeed is unavoidable. Therefore, we can ignore the SM sector in
our later calculations, but only keep the supersymmetric
contributions, namely in the following section, we only consider
the three classes of contributions (Fig.\ref{fig1}(e,f,g)) to the
$\Big(\beta\beta\Big)_{0\nu}$ decay. The explicit Feynman diagrams
of the three classes at one-loop level are drawn in Fig.\ref{fig2}
and Fig.\ref{fig3}. In those figures, Fig.\ref{fig2}(a,b) belong
to the class shown in Fig.\ref{fig1}(e); Fig.\ref{fig2}(c,d,e) and
Fig.\ref{fig3}(h,i) belong to the class shown in Fig.\ref{fig1}(f)
and Fig.\ref{fig2}(f,g) and Fig.\ref{fig3}(j,k,l,m) belong to the
class shown in Fig.\ref{fig1}(g).

After integrating out those heavy internal particles, the
effective Lagrangian for $dd\rightarrow uu+e^-e^-$ is written as
\begin{eqnarray}
&&{\cal L}_{_{eff}}^{^{\Delta L_e=2}}=\frac{G_F^2}{2m_P}\bigg\{\bigg[
C_1(\mu_{_{\rm W}})\bar{u}_\alpha\gamma_\rho(1-\gamma_5)d_\alpha
\bar{u}_\beta\gamma^\rho(1-\gamma_5)d_\beta\bar{e}(1+
\gamma_5)e^c\nonumber \\
&&\hspace{1.8cm}+C_2(\mu_{_{\rm W}})
\bar{u}_\alpha\gamma_\rho(1-\gamma_5)d_\alpha
\bar{u}_\beta\gamma^\rho(1-\gamma_5)d_\beta\bar{e}(1-
\gamma_5)e^c\nonumber \\
&&\hspace{1.8cm}+C_3(\mu_{_{\rm W}})\bar{u}_\alpha(1+\gamma_5)d_\alpha\bar{u}_\beta
\gamma_\rho(1-\gamma_5)d_\beta\bar{e}\gamma^\rho(1-\gamma_5)
e^c\nonumber \\&&\hspace{1.8cm}
+C_4(\mu_{_{\rm W}})\bar{u}_\alpha
(1-\gamma_5)d_\alpha\bar{u}_\beta\gamma_\rho(1-\gamma_5)d_\beta
\bar{e}\gamma^\rho(1-\gamma_5)e^c\nonumber \\
&&\hspace{1.8cm}
+C_5(\mu_{_{\rm W}})\bar{u}_\alpha
(1+\gamma_5)d_\alpha\bar{u}_\beta\gamma_\rho(1-\gamma_5)d_\beta
\bar{e}\gamma^\rho(1+\gamma_5)e^c\nonumber \\
&&\hspace{1.8cm}+C_6(\mu_{_{\rm W}})\Big(\bar{u}_\alpha
\omega_-d_\alpha\bar{u}_\beta\omega_-d_\beta+{1\over 4}\bar{u}
\sigma_{\mu\nu}\omega_-d_\alpha\bar{u}_\beta\sigma^{\mu\nu}\omega_-
d_\beta\Big)\bar{e}\omega_+e^c\nonumber \\
&&\hspace{1.8cm}+C_7(\mu_{_{\rm W}})\Big(\bar{u}_\alpha
\omega_+d_\alpha\bar{u}_\beta\omega_+d_\beta+{1\over 4}\bar{u}
\sigma_{\mu\nu}\omega_+d_\alpha\bar{u}_\beta\sigma^{\mu\nu}\omega_+
d_\beta\Big)\bar{e}\omega_-e^c\bigg]\nonumber \\
&&\hspace{1.8cm}+\bigg[C_8(\mu_{_{\rm W}})
\bar{u}_\alpha\gamma_\rho(1-\gamma_5)d_\alpha
\bar{u}_\beta\gamma_\sigma(1-\gamma_5)d_\beta
\bar{e}(i\sigma^{\rho\sigma})(1+\gamma_5)e^c\nonumber\\
&&\hspace{1.8cm}+C_9(\mu_{_{\rm W}})
\bar{u}_\alpha\gamma_\rho(1-\gamma_5)d_\alpha
\bar{u}_\beta\gamma_\sigma(1-\gamma_5)d_\beta
\bar{e}(i\sigma^{\rho\sigma})(1-\gamma_5)e^c\nonumber\\
&&\hspace{1.8cm}+C_{10}(\mu_{_{\rm W}})
\bar{u}_\alpha(i\sigma_{\rho\sigma})\omega_+d_\alpha
\bar{u}_\beta\gamma^\rho\omega_-d_\beta\bar{e}\gamma^\sigma
\omega_+e^c\nonumber\\
&&\hspace{1.8cm}+C_{11}(\mu_{_{\rm W}})
\bar{u}_\alpha(i\sigma_{\rho\sigma})\omega_-d_\alpha
\bar{u}_\beta\gamma^\rho\omega_-d_\beta\bar{e}\gamma^\sigma
\omega_-e^c\nonumber\\
&&\hspace{1.8cm}+C_{12}(\mu_{_{\rm W}})
\bar{u}_\alpha(i\sigma_{\rho\sigma})\omega_+d_\alpha
\bar{u}_\beta\gamma^\rho\omega_-d_\beta\bar{e}\gamma^\sigma
\omega_-e^c\nonumber\\
&&\hspace{1.8cm}
-{1\over 8}\eta_{d_1}(\mu_{_{\rm W}})\epsilon_{\mu\rho\sigma\delta}
\bar{u}_\alpha\gamma^\mu\omega_-d_\alpha\bar{u}_\beta\sigma^{\rho\sigma}
\omega_+d_\beta\bar{e}\gamma^\delta\omega_+e^c\nonumber\\
&&\hspace{1.8cm}+C_{13}(\mu_{_{\rm W}})\epsilon_{\mu\rho\sigma\delta}
\bar{u}_\alpha\gamma^\mu\omega_-d_\alpha\bar{u}_\beta\sigma^{\rho\sigma}
\omega_-d_\beta\bar{e}\gamma^\delta\omega_-e^c\nonumber\\
&&\hspace{1.8cm}+C_{14}(\mu_{_{\rm W}})\epsilon_{\mu\rho\sigma\delta}
\bar{u}_\alpha\gamma^\mu\omega_-d_\alpha\bar{u}_\beta\sigma^{\rho\sigma}
\omega_+d_\beta\bar{e}\gamma^\delta\omega_-e^c\nonumber \\
&&\hspace{1.8cm}-{1\over 8}\eta_{e_1}(\mu_{_{\rm W}})
\epsilon_{\mu\rho\sigma\delta}
\bar{u}_\alpha\gamma^\mu\omega_-d_\alpha\bar{u}_\beta\gamma^\nu
\omega_-d_\beta\bar{e}\sigma^{\rho\sigma}\omega_-e^c\nonumber \\
&&\hspace{1.8cm} -{1\over 8}\eta_{e_2}(\mu_{_{\rm
W}})\epsilon_{\mu\rho\sigma\delta}
\bar{u}_\alpha\gamma^\mu\omega_-d_\alpha\bar{u}_\beta\gamma^\nu
\omega_-d_\beta\bar{e}\sigma^{\rho\sigma}\omega_+e^c\bigg]\bigg\}.
\label{lag}
\end{eqnarray}
The explicit expression of $\eta_i(\mu_{_{\rm W}})$ and
$C_j(\mu_{_{\rm W}})$ are given in appendix \ref{appb}. We
integrate out those heavy fields at the electro-weak scale
$\mu_{_{\rm W}}$. Here, we have used the Fierz transformation to
re-arrange the quark currents into color-singlet forms.

The next step of the calculations is to re-formulate the
transition matrix amplitudes where all quark degrees of freedom
are replaced by the corresponding nucleon degrees of freedom in
nuclei and it is based on the principle so-called as "quark-hadron duality". To
carry out these calculations, one needs concrete forms of the
relevant nuclear structure, i.e. the wavefunctions which describe
the nuclei.

\section{Nuclear $\Big(\beta\beta\Big)_{0\nu}$ decay}

At the nuclear level, the amplitude for
$\Big(\beta\beta\Big)_{0\nu}$ is written as
\begin{equation}
\langle(A,Z+2),2e^-|S-1|(A,Z)\rangle =
\langle(A,Z+2),2e^-|Texp\Big[i\int d^4x{\cal L}_{_{eff}}^
{^{\Delta L_e=2}}\Big]|(A,Z)\rangle\;,
\label{amp-nuc}
\end{equation}
where the effective ${\cal L}_{_{eff}}^{^{\Delta L_e=2}}$ is given
in Eq.(\ref{lag}). The nuclear structure is involved via the
initial $(A,Z)$ and the final $(A,Z+2)$ nuclear states which have
the same atomic weight $A$, but different electric charges $Z$ and
$Z+2$. The standard framework for the calculation of the nuclear
matrix elements is the nonrelativistic impulse approximation
(NRIA)\cite{Doi}. Now, we turn to the derivation of the nuclear
matrix elements of the color singlet currents in Eq.(\ref{lag})
using the results of Ref.\cite{Adler}. The relevant matrix
elements of those currents are
\begin{eqnarray}
&&\langle P(p)|\bar{u}d|N(p')\rangle=F_S^{(3)}(q^2)\bar{{\cal N}}(p)\tau_+
{\cal N}(p')\;,\nonumber \\
&&\langle P(p)|\bar{u}\gamma_5d|N(p')\rangle=F_P^{(3)}(q^2)\bar{{\cal N}}(p)
\gamma_5\tau_+{\cal N}(p')\;,\nonumber \\
&&\langle P(p)|\bar{u}\gamma^\rho(1-\gamma_5)d|N(p')\rangle=\bar{{\cal N}}(p)
\Big(F_V(q^2)-F_A(q^2)\gamma_5+F_W(q^2)i\sigma^{\rho\sigma}q_{\sigma}
+F_P(q^2)q_\rho\gamma_5\Big)\tau_+{\cal N}(p')\;,\nonumber \\
&&\langle P(p)|\bar{u}\gamma^\rho(1+\gamma_5)d|N(p')\rangle=\bar{{\cal N}}(p)
\Big(F_V(q^2)+F_A(q^2)\gamma_5+F_W(q^2)i\sigma^{\rho\sigma}q_{\sigma}
-F_P(q^2)q_\rho\gamma_5\Big)\tau_+{\cal N}(p')\;,
\nonumber \\
&&\langle P(p)|\bar{u}\sigma^{\rho\sigma}(1+\gamma_5)|N(p')\rangle
=\bar{\cal N}(p)\Big(J^{\rho\sigma}+{i\over 2}\epsilon^{\rho\sigma
\lambda\delta}J_{\lambda\delta}\Big)\tau_+{\cal N}(p')\;,
\nonumber \\
&&\langle P(p)|\bar{u}\sigma^{\rho\sigma}(1-\gamma_5)|N(p')\rangle
=\bar{\cal N}(p)\Big(J^{\rho\sigma}-{i\over 2}\epsilon^{\rho\sigma
\lambda\delta}J_{\lambda\delta}\Big)\tau_+{\cal N}(p')\;,
\label{current}
\end{eqnarray}
where ${\cal N}=\left(\begin{array}{c}P\\N\end{array}\right)$ is a
nucleon isodoublet, and $q=p-p'$. The tensor structure is defined
as
\begin{eqnarray}
&&J^{\mu\nu}=T_1^{(3)}(q^2)\sigma^{\mu\nu}+\frac{iT_2^{(3)}(q^2)}
{m_P}(\gamma^\mu q^\mu-\gamma^\nu q^\mu)+\frac{T_3^{(3)}(q^2)}
{m_P^2}(\sigma^{\mu\rho}q_\rho q^\nu-\sigma^{\nu\rho}q_\rho
q^\mu)\;.
\label{tensor}
\end{eqnarray}

For all form factors, we take the dipole form\cite{Vergados3}
\begin{equation}
\frac{F_{V,A}(q^2)}{f_{V,A}}=\frac{F_{S,P,W}(q^2)}{F_{S,P,W}(0)}
=\frac{T_i^{(3)}(q^2)}{T_i^{(3)}(0)}=(1-\frac{q^2}{m_A^2})^{-2}
\end{equation}
with $m_A=0.85{\rm GeV}$ and $f_V\approx 1$, $f_A\approx 1.261$.
The nuclear couplings obey the relations
\begin{eqnarray}
&&{F_{W}(0)\over f_V}=-\frac{\mu_p-\mu_n}{2m_P}\approx {3.7\over 2m_P}
\nonumber \\
&&{F_P(0)\over f_A}={2m_P\over m_\pi^2}\;,
\label{coup-rela}
\end{eqnarray}
where $m_{P(\pi)}$ is the proton (pion) mass and $\mu_{p(n)}$ is
the proton (neutron) magnetic moment. With the nonrelativistic
quark model,
$T_1^{(3)}(0)=1.45,\;T_2^{(3)}=-1.48,\;T_3^{(3)}=-0.66$\cite{Adler}.

The transition operators are defined as
\begin{eqnarray}
&&\Omega_{F,N}=\sum\limits_{a\neq b}\tau_a^+\tau_b^+
\bigg({R_0\over r_{ab}}\bigg)F_N(x_A)\;,\nonumber \\
&&\Omega_{GT,N}=\sum\limits_{a\neq b}\tau_a^+\tau_b^+
{\bf \sigma}_a\cdot{\bf \sigma}_b
\bigg({R_0\over r_{ab}}\bigg)F_N(x_A)\;\nonumber\\
&&\Omega_{GT'}=\sum\limits_{a\neq b}\tau_a^+\tau_b^+
{\bf \sigma}_a\cdot{\bf \sigma}_b
\bigg({R_0\over r_{ab}}\bigg)F_4(x_A)\;,\nonumber \\
&&\Omega_{T'}=\sum\limits_{a\neq b}\tau_a^+\tau_b^+
\Big\{3({\bf \sigma}_a\cdot {\bf \hat{r}}_{ab})
({\bf \sigma}_b\cdot {\bf \hat{r}}_{ab})-{\bf\sigma}_a
\cdot{\bf\sigma}_b\Big\}
\bigg({R_0\over r_{ab}}\bigg)F_5(x_A)\;,\nonumber \\
&&{\bf\Omega}_{V,N}=\sum\limits_{a\neq b}\tau_a^+\tau_b^+
{\mathbf \sigma}_b
\bigg({R_0\over r_{ab}}\bigg)F_N(x_A)\;,\nonumber \\
&&{\bf\Omega}_{V,1}=\sum\limits_{a\neq b}\tau_a^+\tau_b^+
{\mathbf \sigma}_b\times{\bf\hat{r}}_{ab}
\bigg({R_0\over r_{ab}}\bigg)F_6(x_A)\;,\nonumber \\
&&{\bf\Omega}_{V,2}=\sum\limits_{a\neq b}\tau_a^+\tau_b^+
{\mathbf \sigma}_a
\bigg({R_0\over r_{ab}}\bigg)F_4(x_A)\;,\nonumber \\
&&{\bf\Omega}_{V,3}=\sum\limits_{a\neq b}\tau_a^+\tau_b^+
\Big\{3({\bf\sigma}_a\cdot{\bf\hat{r}}_{ab}){\bf\hat{r}}_{ab}
\Big\}\bigg({R_0\over r_{ab}}\bigg)F_5(x_A)\;.
\label{trano2}
\end{eqnarray}
Here, $R_0$ is the nuclear radius being introduced to make the
matrix elements dimensionless and other notations are:
\begin{equation}
{\bf r}_{ab}=({\bf r}_a-{\bf r}_b),\;\;r_{ab}=|{\bf r}_{ab}|
,\;\;{\bf \hat{r}}_{ab}={{\bf r}_{ab}\over r_{ab}},\;\;
x_A=m_Ar_{ab}\;,
\label{notation}
\end{equation}
where ${\bf r}_i$ is the coordinate of the {\it i-}th nucleon. The
above matrix elements have been written in the closure
approximation, which is well satisfied due to the large masses of
the inter-loop particles. The expressions for those structure
functions $F_i$ are given \cite{Hirsch}
\begin{eqnarray}
&&F_N(x)=\frac{x}{48}(3+3x+x^2)e^{-x}\;,\nonumber \\
&&F_4(x)=\frac{x}{48}(3+3x-x^2)e^{-x}\;,\nonumber \\
&&F_5(x)=\frac{x^3}{48}e^{-x}\;,\nonumber \\
&&F_6(x)=\frac{x}{48}(1+x)e^{-x}\;.\nonumber \\
\label{Fi}
\end{eqnarray}
With these form factors, the reaction matrix element is obtained
as
\begin{eqnarray}
&&{\cal R}_{\Big(\beta\beta\Big)_{0\nu}}^{0^+\rightarrow
0^+}=\frac{G_F^2}{\sqrt{2}m_PC_{0\nu}}
\bigg\{C_1(\mu_{_{\rm W}})\Omega_{V-A}\Big[\bar{e}(1+\gamma_5)e^c\Big]
+C_2(\mu_{_{\rm W}})
\Omega_{V-A}\Big[\bar{e}(1-\gamma_5)e^c\Big]\nonumber\\
&&\hspace{1.8cm}
+C_3(\mu_{_{\rm W}})\Omega_{1}^0\Big[\bar{e}\gamma_0(1-\gamma_5)
e^c\Big]+C_4(\mu_{_{\rm W}})\Omega_{2}^0
\Big[\bar{e}\gamma_0(1-\gamma_5)e^c\Big]\nonumber \\
&&\hspace{1.8cm}
+C_5(\mu_{_{\rm W}})\Omega_{1}^0\Big[\bar{e}\gamma_0(1+\gamma_5)e^c\Big]
+C_6(\mu_{_{\rm W}})\Omega_{ST}\Big[\bar{e}(1+\gamma_5)e^c\Big]\nonumber \\
&&\hspace{1.8cm}+C_7(\mu_{_{\rm W}})
\Omega_{ST}\Big[\bar{e}(1-\gamma_5)e^c\Big]\bigg\}\;, \label{reac}
\end{eqnarray}
where
\begin{eqnarray}
&&\Omega_1^\rho=\frac{m_P}{m_e}\bigg\{g^{\rho 0}\Big[
\frac{F_S^{(3)}(0)f_V}{f_A^2}\Omega_{F,N}-{1\over 12}
\bigg(\frac{m_A}{m_P}\bigg)^2\frac{F_S^{(3)}(0)}{f_A}
\Big(\Omega_{GT'}-\Omega_{T'}\Big)\Big]+g^{\rho k}\Big[
\frac{F_S^{(3)}(0)}{f_A}\Omega_{V,N}^k\nonumber \\
&&\hspace{1.1cm}-\frac{m_A}{2m_P}\;\frac{F_S^{(3)}(0)f_V}
{f_A^2}\Omega_{V_1}^k-\frac{1}{12}\bigg({m_A\over m_P}\bigg)^2
\frac{F_P^{(3)}(0)f_V}{f_A^2}\Big(\Omega_{V_2}^k-\Omega_{V_3}^k\Big)
\Big]\bigg\}\;,\nonumber \\
&&\Omega_2^\rho=\Omega_1^\rho(F_P^{(3)}(0)\rightarrow -F_P^{(3)}(0))\;,
\nonumber\\
&&\Omega_{V-A}=\frac{m_{P}}{m_e}\bigg\{\Big(\frac{f_V}{f_A}\Big)^2
\langle F|\Omega_{F,N}|I\rangle-\langle
F|\Omega_{GT,N}|I\rangle\bigg\}\;,\nonumber \\
&&\Omega_{ST}=\frac{m_{P}}{m_e}\bigg\{\Big({F_S^{(3)}(0)\over
f_A}\Big)^2\langle F|\Omega_{F,N}|I\rangle -\Big({T_1^{(3)}(0)\over
f_A}\Big)^2\langle F|\Omega_{GT,N}|I\rangle +\Big({m_A\over m_P}\Big)^2
\Big[{1\over 4}\Big({T_1^{(3)}(0)\over f_A}\Big)^2\nonumber \\
&&\hspace{1.2cm}+\Big({T_2^{(3)}(0)\over
f_A}\Big)^2-\frac{T_1^{(3)}(0)
T_2^{(3)}(0)}{f_A^2}\Big]\Omega_{F'}+\Big({m_A\over m_P}\Big)^2\Big[{1\over 6}
\Big({T_1^{(3)}(0)\over f_A}\Big)^2-{2\over 3}\frac{T_1^{(3)}(0)T_2^{(3)}(0)}
{f_A^2}\nonumber \\
&&\hspace{1.2cm}+{4\over 3}\frac{T_1^{(3)}(0)T_3^{(3)}(0)}{f_A^2}
-{1\over 12}\Big(\frac{F_P^{(3)}(0)}{f_A}\Big)^2\Big]\Omega_{GT'}
+\Big({m_A\over m_P}\Big)^2\Big[{1\over 12}
\Big({T_1^{(3)}(0)\over f_A}\Big)^2-{1\over 3}\frac{T_1^{(3)}(0)T_2^{(3)}(0)}
{f_A^2}\nonumber \\
&&\hspace{1.2cm}+{2\over 3}\frac{T_1^{(3)}(0)T_3^{(3)}(0)}{f_A^2}
+{1\over
12}\Big(\frac{F_P^{(3)}(0)}{f_A}\Big)^2\Big]\Omega_{T'}\bigg\}\;.
\label{v-a}
\end{eqnarray}
The $0^+\rightarrow 0^+$ decay rate $d\Gamma_{0\nu\beta\beta}
^{0^+\rightarrow 0^+}$ for the process $N_i(A,Z-2)\rightarrow
N_f(A,Z)+e_1+e_2$ is written as
\begin{eqnarray}
&&d\Gamma_{0\nu\beta\beta}^{0^+\rightarrow 0^+}=2\pi\sum
\limits_{spin}|{\cal R}_{0\nu\beta\beta}^{0^+\rightarrow 0^+}|^2
\delta(\epsilon_1+\epsilon_2+E_f-M_i)d\Omega_{e_1}d\Omega_{e_2}
\nonumber\\
&&\hspace{1.6cm}=\frac{G_F^4}{32\pi^5}\frac{(f_Am_A)^4}{(R_0m_P)^2}
\bigg\{A_0^{0\nu\beta\beta}+{\bf \hat{p}}_1\cdot{\bf \hat{p}}_2
B_0^{0\nu\beta\beta}+\Big[\Big({\bf \hat{p}}_1\cdot{\bf \hat{p}}_2
\Big)^2-\frac{1}{3}\Big]C_0^{0\nu\beta\beta}\bigg\}\nonumber \\
&&\hspace{2cm}p_1p_2\epsilon_1
\epsilon_2\delta(\epsilon_1+\epsilon_2+E_f-M_i)d\epsilon_1d\epsilon_2
d({\bf \hat{p}}_1\cdot{\bf \hat{p}}_2)
\label{drate}
\end{eqnarray}
with
\begin{eqnarray}
&&A_0^{(\beta\beta)_{0\nu}}=\Big|\tilde{g}_{-1}(\epsilon_1)
\tilde{g}_{-1}(\epsilon_2)\Big|^2\Big|\Big(C_1(\mu_{_{\rm W}})
-C_2(\mu_{_{\rm W}})\Big)\Omega_{V-A}
+\Big(C_3(\mu_{_{\rm W}})+C_5(\mu_{_{\rm W}})\Big)\Omega_{1}^0
+C_4(\mu_{_{\rm W}})\Omega_{2}^0\nonumber \\
&&\hspace{1.5cm}
+\Big(C_6(\mu_{_{\rm W}})-C_7(\mu_{_{\rm W}})\Big)\Omega_{ST}\Big|^2
+\frac{1}{3}\Big|\tilde{f}_{1}(\epsilon_1)
\tilde{f}_{1}(\epsilon_2)\Big|^2\Big|\Big(C_1(\mu_{_{\rm W}})
-C_2(\mu_{_{\rm W}})\Big)\Omega_{V-A}
\nonumber \\
&&\hspace{1.5cm}
-\Big(C_3(\mu_{_{\rm W}})+C_5(\mu_{_{\rm W}})\Big)\Omega_{1}^0
-C_4(\mu_{_{\rm W}})\Omega_{2}^0
+\Big(C_6(\mu_{_{\rm W}})-C_7(\mu_{_{\rm W}})
\Big)\Omega_{ST}\Big|^2
\;,\nonumber\\
&&B_0^{(\beta\beta)_{0\nu}}=2\tilde{f}_{1}(\epsilon_1)
\tilde{f}_{1}(\epsilon_2)\tilde{g}_{-1}(\epsilon_1)
\tilde{g}_{-1}(\epsilon_2)\bigg\{\Big|\Big(C_3(\mu_{_{\rm W}})
+C_5(\mu_{_{\rm W}})\Big)\Omega_{1}^0
+C_4(\mu_{_{\rm W}})\Omega_{2}^0\Big|^2
\nonumber\\&&\hspace{1.5cm}
-\Big|\Big(C_1(\mu_{_{\rm W}})-C_2(\mu_{_{\rm W}})\Big)
\Omega_{V-A}+\Big(C_6(\mu_{_{\rm W}})-C_7(\mu_{_{\rm W}})
\Big)\Omega_{ST}\Big|^2\bigg\}\;,\nonumber\\
&&C_0^{(\beta\beta)_{0\nu}}=\Big|\tilde{f}_{1}(\epsilon_1)
\tilde{f}_{1}(\epsilon_2)\Big|^2\Big|\Big(C_1(\mu_{_{\rm W}})
-C_2(\mu_{_{\rm W}})\Big)
\Omega_{V-A}-\Big(C_3(\mu_{_{\rm W}})+C_5(\mu_{_{\rm W}})
\Big)\Omega_{1}^0\nonumber \\
&&\hspace{1.5cm} -C_4(\mu_{_{\rm W}})\Omega_{2}^0
+\Big(C_6(\mu_{_{\rm W}})-C_7(\mu_{_{\rm W}})
\Big)\Omega_{ST}\Big|^2\;, \label{abc}
\end{eqnarray}
where the expressions of $\tilde{f},\;\tilde{g}$ are taken from
Ref.\cite{Doi}.
\section{The input parameters and numerical analysis}

In this section, we  present our numerical analysis on the
neutrinoless double beta decay in the supersymmetric seesaw
models. In order to simplify our discussion, we assume that the
Yukawa interaction and soft-breaking part of the Lagrangian all
are flavor-conserved, the flavor changing interactions are
mediated by the CKM entries of the quark and lepton sectors. This
is just the so called 'minimal flavor violation' scenario in the
supersymmetric models. Under the assumption, there are four
couplings, $g_1,\;g_2,\;g_3$ and $\mu$, the right-handed neutrino
masses $m_{_I}^n$, the Yukawa couplings for neutrinos
$h_{_I}^{\nu}\; (I=1,\cdots,N_{_G})$ and $6+11N_{_{G}}$ free
independent parameters for supersymmetry breaking, all together
$10+13N_{_G}$ parameters to be fixed besides the lepton CKM matrix
elements. Among them, 11 parameters for each generation appear
only in the sfermion mass matrix and the mixing. Therefore, it is
more convenient to choose eight physical masses and three mixing
angles for the three charged sfermions in each generation as input
parameters. As for the right-handed neutrino parameters, we also
set $m_1^n=m_2^n=\cdots =m_{_{N_G}}^n=m_{_R}$ and the matrix ${\bf
U_{_{\rm M}}}={\bf I}$ for simplicity. The leptonic sector CKM
matrix simply is ${\bf U}$ in this case. For each generation, we
can specify the relevant input parameters as three scalar quark
masses
$m_{_{\tilde{U}_1^I}},\;m_{_{\tilde{U}_2^I}},\;m_{_{\tilde{D}_1^I}}$,
two mixing angles
$\theta_{_{\tilde{U}^I}},\;\theta_{_{\tilde{D}^I}}$ for the squark sector,
two-charged-scalar lepton masses and one mixing angle
$m_{_{\tilde{E}_1^I}},\;m_{_{\tilde{E}_2^I}},\;\theta_{_{\tilde{E}^I}}$,
three sneutrino masses $m_{_{\tilde{\nu}_1^I}},\;
m_{_{\tilde{\nu}_2^I}},\;m_{_{\tilde{\nu}_3^I}}$ for the slepton sector. As well, the
Yukawa couplings of neutrinos $h_{I}^{\nu}$ are also input
parameters. Another scalar down-type quark mass is obtained
through the relation
\begin{equation}
\cos^2\theta_{_{\tilde{U}_1^I}}m^2_{_{\tilde{U}_1^I}}
+\sin^2\theta_{_{\tilde{U}_1^I}}m^2_{_{\tilde{U}_2^I}}-
m_{_{u^I}}^2=\cos^2\theta_{_{\tilde{D}_1^I}}m^2_{_{\tilde{D}_1^I}}
+\sin^2\theta_{_{\tilde{D}_2^I}}m^2_{_{\tilde{U}_2^I}}-m_{_{d^I}}^2
+m_{_{\rm W}}^2\cos 2\beta\;.
\label{msd2}
\end{equation}
Assuming the relations $m_{_{\tilde{\nu}_1^I}}\ll
m_{_{\tilde{\nu}_2^I}}< m_{_{\tilde{\nu}_3^I}}$ hold among the
three scalar neutrino masses, the sneutrino mixing matrix is
written as (accurate to order ${\cal
O}(({m_{_{\tilde{\nu}_1^I}}\over m_{_{\tilde{\nu}_2^I}}})^2)$)
\begin{eqnarray}
&&Z_{\tilde{\nu}^I}=\left(\begin{array}{ccc}
1&\frac{\sqrt{2}\rho_{_I}}{m_{_{\tilde{\nu}_2^I}}^2}&0\\
-\frac{\rho_{_I}}{m_{_{\tilde{\nu}_2^I}}^2}&{1\over \sqrt{2}}&
-{1\over\sqrt{2}}\\
-\frac{\rho_{_I}}{m_{_{\tilde{\nu}_2^I}}
^2}&{1\over\sqrt{2}}&
{1\over\sqrt{2}}
\end{array}\right)\;,
\label{snumix}
\end{eqnarray}
where
\begin{equation}
\rho_{_I}=\frac{\sqrt{(\zeta_I-m_{_{\tilde{\nu}_1^I}}^2)(
m_{_{\tilde{\nu}_1^I}}^2+m_{_{\tilde{\nu}_2^I}}^2-\zeta_I)}}{\sqrt{2}}
\label{eps}
\end{equation}
with $\zeta_I=\cos^2\theta_{_{\tilde{E}^I}}
m_{_{\tilde{E}_1^I}}^2+\sin^2\theta_{_{\tilde{E}^I}}
m_{_{\tilde{E}_2^I}}^2-m_{_{e^I}}^2-m_{_{\rm W}}^2
(C_{_\beta}^2-S_{_\beta}^2)$. A point needs to be specified that
${\bf B}^n_I\;(I=1,\;2,\;\cdots,\;N_{_G})$ are negative when we
derived  Eq.\ref{snumix}.
For the neutrinoless double $\beta$ decay, the lepton number is
violated, thus only the Majorana component of the neutrino or
scalar neutrino is responsible for the process. Therefore the
amplitude must be proportional to the mixing entry.
Diagonalizing the mass-square matrix, there are three
eigen-values, which are
$m_{_{\tilde{\nu}_1^I}}^2=\xi_{_I}-\frac{2\rho_{_I}^2}
{m_{_{\tilde{\nu}_2^I}}^2}\;,$ $m_{_{\tilde{\nu}_2^I}}^2={1\over
2}\bigg( m_{_{n^I}}^2+2h_{_I}^{\nu^2}\upsilon_2^2 +m_{_{N^{I}}}^2+
B^n_{_I}\bigg)\;,$ $ m_{_{\tilde{\nu}_3^I}}^2={1\over 4}\bigg(
m_{_{n^I}}^2+2h_{_I}^{\nu^2}\upsilon_2^2 +m_{_{N^{I}}}^2-
B^n_{_I}\bigg)$, respectively and here
$m_{_{n^I}}^2+2h_{_I}^{\nu^2}\upsilon_2^2 +m_{_{N^{I}}}^2$ is a
positive constant. The corresponding eigenvector of
$m_{_{\tilde{\nu}_1^I}}^2$ is
\begin{equation}
\tilde{\nu}_1^I=\tilde L_1^I-{2\rho_I\over
m_{_{n^I}}^2+2h_{_I}^{\nu^2}\upsilon_2^2 +m_{_{N^{I}}}^2+{\bf
B}_I^n}\tilde N^I-{2\rho_I\over
m_{_{n^I}}^2+2h_{_I}^{\nu^2}\upsilon_2^2 +m_{_{N^{I}}}^2+{\bf
B}_I^n}\tilde N^{I*},
\end{equation}
where $\tilde L_1^I$, $\tilde N^I$ and $\tilde N^{I*}$ are one
Dirac component and two Majorana components in the mass
eigenvector. The mixing entry is proportional to ${2\rho_{_I}\over
m_{_{n^I}}^2+2h_{_I}^{\nu^2}\upsilon_2^2 +m_{_{N^{I}}}^2+{\bf
B}_I^n}$. If ${\bf B_I^n}$ is negative, obviously the mixing is
larger (in this case $m_{\tilde{\nu}_1^I}<m_{\tilde{\nu}_2^I}
<m_{\tilde{\nu}_3^I}$) and the total amplitude is proportional to
$({2\rho_I\over m_{_{n^I}}^2+2h_{_I}^{\nu^2}\upsilon_2^2
+m_{_{N^{I}}}^2+{\bf B}_I^n})^2\sim {m_{_{\tilde{E}^I}}\over
m_{\tilde{\nu}_2^I}}$. Otherwise if  ${\bf B}_I^n$ is positive the
amplitude is still proportional to ${m_{_{\tilde{E}^I}}\over
m_{\tilde{\nu}_2^I}}$, but the relation becomes
$m_{\tilde{\nu}_1^I}<m_{\tilde{\nu}_3^I}<m_{\tilde{\nu}_2^I}$
instead, and the mixing would be relatively small. To meet the
data, the mixing cannot be too small. In the three generation
neutrino case, the contribution is suppressed by the factor
$<m_{\nu}>/ p_F\sim 10^{-10}$ in addition to the electro-weak
coupling factor, so that cannot be substantial to explain the
double beta decay data. In this SUSY model with right-handed
neutrinos, the mixing factor is ${m_{_{\tilde{E}^I}}\over
m_{\tilde{\nu}_2^I}}$, typically, $m_{_{\tilde{E}^I}}$ is about
several TeV, so if $m_{\tilde{\nu}_2}$ is of a medium energy
scale, (about $10^{7}$ GeV), its contribution can meet the
observed data on the double beta decay.

Concerning the remaining relevant parameters,
$\mu,\;m_1,\;m_2,\;m_3$, it is customary to use, in place of
$\mu,\;m_2,m_3$, the two chargino masses $m_{_{\kappa^+_{1,2}}}$,
gluino mass $m_{_{\tilde{g}}}$ and $m_1$. From  Eq.\ref{spot1} and
Eq.\ref{soft}, one can easily express the original parameters
appearing in the Lagrangian in terms of the physical input
parameters (the physical masses and mixing angles of sfermions).
For the charged sfermion sector, the original parameters are
expressed in terms of the  physical input parameters and the
expressions can be found in Ref.\cite{notation}. We present the
explicit expressions of the original parameters of the sneutrino
sector in terms of the physical input parameters in appendix
\ref{appa}. The consequent new relevant physical parameters are:
$\tan\beta,\;m_1,\;m_{_{\tilde{g}}},\;m_{_{\kappa_{1,2}^+}},\;m_{_{\tilde{U}_1^1}},\;
m_{_{\tilde{U}_2^1}},\;m_{_{\tilde{D}_1^1}},\;
\theta_{_{\tilde{U}^1}},\;\theta_{_{\tilde{D}^1}},\;
m_{_{\tilde{E}_1^I}},\;m_{_{\tilde{E}_2^I}},\;\theta_{_{\tilde{E}^I}},
\;m_{_{\tilde{\nu}_1^I}},\;m_{_{\tilde{\nu}_2^I}},
\;m_{_{\tilde{\nu}_3^I}},$ $h_{I}^\nu,\;m_{_R},\;{\bf U}_{3\times
3} \;(I=1,\;2,\;3)$ plus the SM input parameters.

As for the SM parameters, we take $m_\tau=1.78{\rm GeV},\;
m_b=5{\rm GeV},\; m_t=174{\rm GeV},\; m_{\rm Z}=91.18{\rm GeV} ,\;
m_{\rm W}=80.33{\rm GeV},\;\alpha_e(m_{\rm W})=\frac{1}{128},\;
\alpha_s(m_{\rm W})=0.12$ at the weak scale. Choosing the mass of
the right-handed neutrino masses is a bit tricky. In order to
explain the data of the solar and atmospheric neutrino
experiments, the effective neutrino Majorana mass
$m_{_{\nu_M}}<0.01{\rm eV}$ is required in the most favored
neutrino mixing scenarios. When setting $m_{\nu}\sim 0.01$ eV, if
assuming the Yukawa coupling of neutrinos $h_{\nu}$ is
approximately equal to unity, the seesaw mechanism requires the
mass of the right-handed neutrino to be $m_{_R}\sim 10^{14}$ GeV,
if $h_{\nu}\approx 0.1$, we have $m_{_R}\sim 10^{12}$ GeV. For we
take three right-handed neutrinos with the same mass $m_{_R}$, we
assume that the neutrino Yukawa couplings have the relation
$h_3^\nu\gg h_{1,2}^\nu$ in order to fit the data of the
atmospheric neutrino and solar neutrino experiments.

In our  calculations, we take the input supersymmetric physical
parameters as following
$$m_{_{\tilde{U}^1_1}}=m_{_{\tilde{U}^1_2}}=m_{_{\tilde{D}^1_1}}=2{\rm TeV},
\;\theta_{_{\tilde{U}^1}}=\theta_{_{\tilde{D}^1}}={\pi\over 2},$$
$$m_{_{\tilde{E}^1_1}}=m_{_{\tilde{E}^1_2}}=m_{_{\tilde{E}^2_1}}=
m_{_{\tilde{E}^2_2}}=m_{_{\tilde{E}^3_2}}=2{\rm TeV},\;
m_{_{\tilde{E}^3_1}}=200{\rm GeV},\;\theta_{_{\tilde{E}^I}}
={\pi\over 2}\;(I=1,\;2,\;3),$$
$$m_{_{\tilde{\nu}^1_1}}=m_{_{\tilde{\nu}^2_1}}=2{\rm TeV},\;
m_{_{\tilde{\nu}^1_2}}=m_{_{\tilde{\nu}^2_2}}=10^8{\rm GeV},\;
m_{_{\tilde{\nu}^1_3}}=m_{_{\tilde{\nu}^2_3}}=m_{_{\tilde{\nu}^3_3}}
=10^{14}{\rm GeV},$$
$$h_1^\nu=0,\;h_2^\nu={h_3^\nu\over 10}\;,\;m_1=m_{_{\tilde{g}}}=300{\rm GeV},\;
m_{_{\kappa_1^+}}=200{\rm GeV},\;m_{_{\kappa_2^+}}=500{\rm GeV}.$$

For the mixing matrix of the lepton sector, we consider two
possibilities,  they are given as
\begin{equation}
{\bf U}_1=\left(\begin{array}{ccc}
0.91&0.35&0.24\\ -0.42&0.72&0.55\\0&-0.60&0.80\end{array}
\right),\;\;\;\;
{\bf U}_2=\left(\begin{array}{ccc}
{1\over \sqrt{2}}&0.5&0.5\\ -{1\over \sqrt{2}}&0.5&0.5\\
0&-{1\over \sqrt{2}}&{1\over\sqrt{2}}\end{array}
\right).
\label{mix1}
\end{equation}
The first mixing matrix of Eq.\ref{mix1} corresponds to the
solution of the solar neutrino anomaly based on the MSW mechanism
with larger mixing angles (LMA) and the second corresponds to the
vacuum oscillation solution for very low values of the squared
mass difference(Just So)\cite{Bilenky}.

\begin{figure}
\setlength{\unitlength}{1mm}
\begin{center}
\begin{picture}(230,200)(55,90)
\put(50,50){\includegraphics{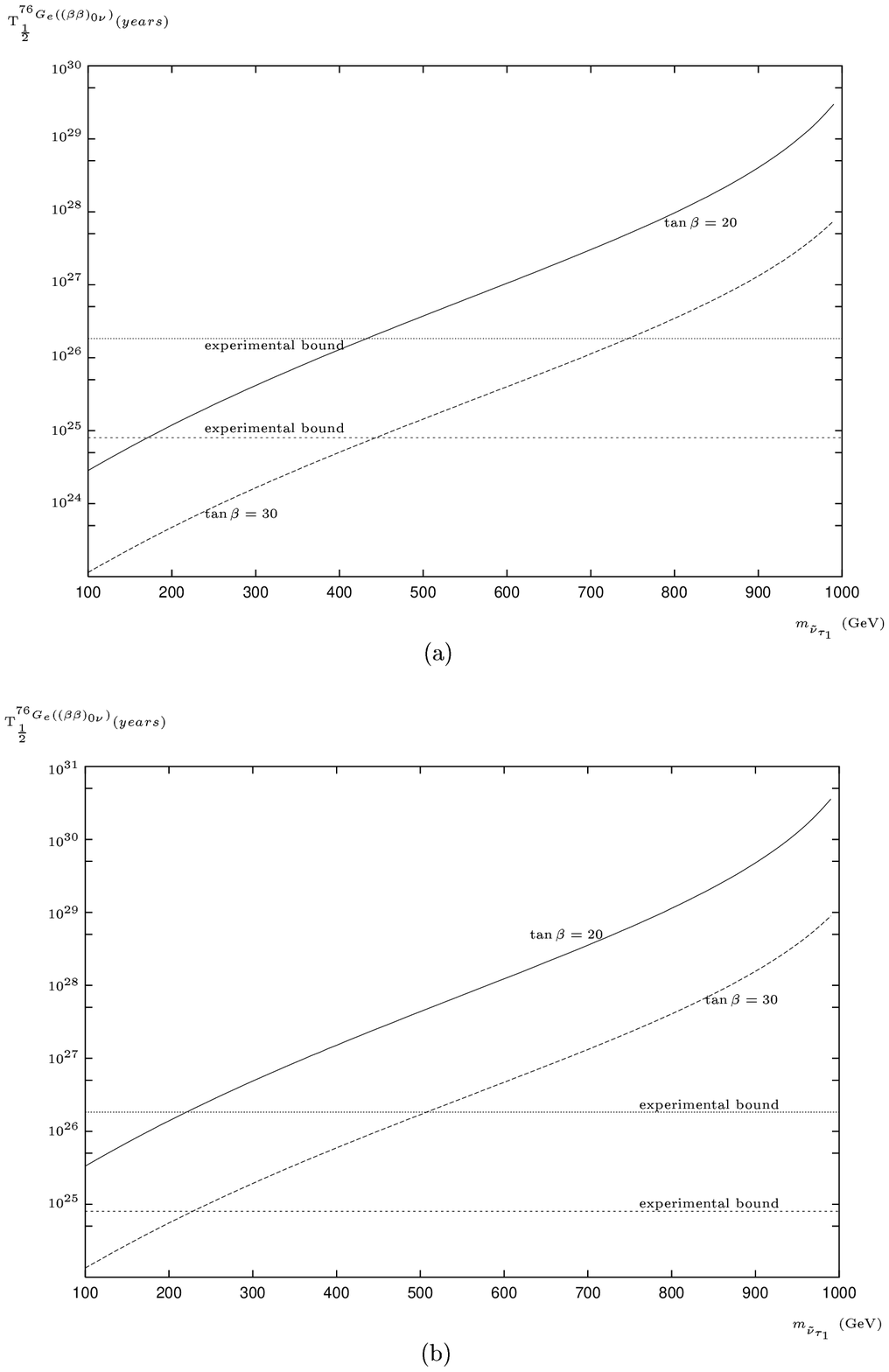}}
\end{picture}
\caption[]{The lifetime of neutrinoless double beta decay
of nuclei $^{76}G_e$ versus the lightest $\tau$-sneutrino
mass with leptonic CKM matrix ${\bf U}_M={\bf U}_1$ and
(a)$h_3^\nu=1,\;m_{_R}=10^{14}{\rm GeV},\;m_{_{\tilde{\nu}_2^3}}
=4\times 10^7{\rm GeV}$; (b)$h_3^\nu=0.1,\;m_{_R}=10^{12}{\rm GeV},\;
m_{_{\tilde{\nu}_2^3}}=4\times 10^7{\rm GeV}$. The solid
line corresponds to $\tan\beta=20$ and the dash line to $\tan\beta=30$;
the other parameters are taken as in the text.}
\label{fig4}
\end{center}
\end{figure}

\begin{figure}
\setlength{\unitlength}{1mm}
\begin{center}
\begin{picture}(230,200)(55,90)
\put(50,50){\includegraphics{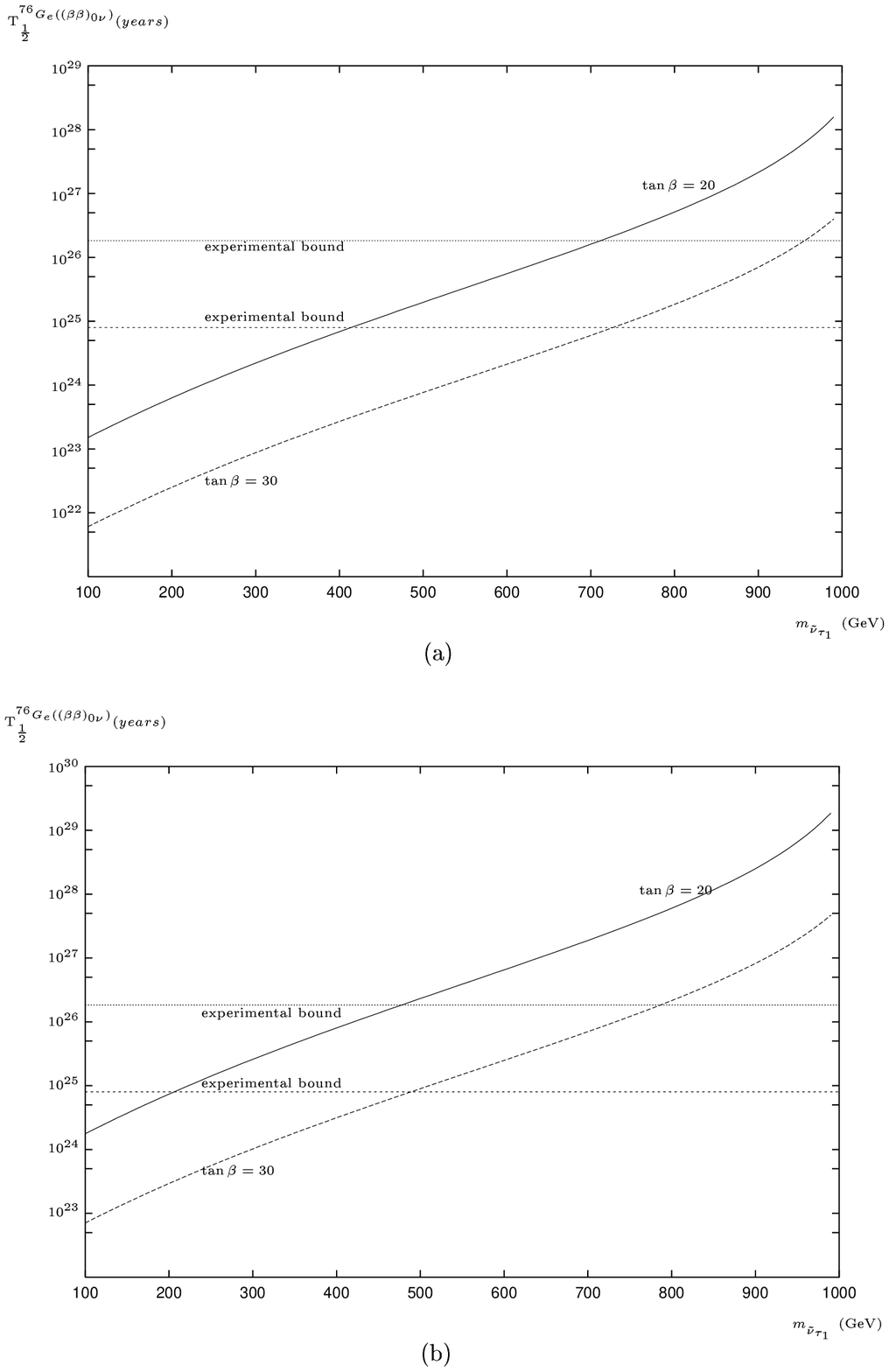}}
\end{picture}
\caption[]{The lifetime of neutrinoless double beta decay
of nuclei $^{76}G_e$ versus the lightest $\tau$-sneutrino
mass with leptonic CKM matrix ${\bf U}_M={\bf U}_2$ and
(a)$h_3^\nu=1,\;m_{_R}=10^{14}{\rm GeV},\;m_{_{\tilde{\nu}_2^3}}
=4\times 10^7{\rm GeV}$; (b)$h_3^\nu=0.1,\;m_{_R}=10^{12}{\rm GeV},\;
m_{_{\tilde{\nu}_2^3}}=4\times 10^7{\rm GeV}$. The solid
line corresponds to $\tan\beta=20$ and the dash line to $\tan\beta=30$;
the other parameters are taken as in the text.}
\label{fig5}
\end{center}
\end{figure}

In order to investigate how the HEIDELBERG-MOSCOW experimental
data constrain the parameter space of the supersymmetric seesaw
model, we compute the lifetime of neutrinoless double beta decay
in nuclei $^{76}G_e$. In Fig.\ref{fig4}, we plot the neutrinoless
double beta decay lifetime of nuclei $^{76}G_e$ versus the
lightest $\tau$-sneutrino mass $m_{_{\tilde{\nu}_1^3}}$. For the
CKM matrix of the lepton sector, we take ${\bf U}_{M}={\bf U}_1$.
Fig.\ref{fig4}(a) corresponds to $h_3^\nu=1.,\;m_{_R}=10^{14}{\rm
GeV}$, and Fig.\ref{fig4}(b) to $h_3^\nu=0.1,\;m_{_R}=10^{12}{\rm
GeV}$. The other relevant parameters are:
$m_{_{\tilde{\nu}_2^3}}=4\times 10^7{\rm GeV}, \;\tan\beta=20$
(solid lines) or $\tan\beta=30$ (dashed-lines). From
Fig.\ref{fig4}, we find that the theoretical prediction on the
$(\beta\beta)_{0\nu}$ decay half life time of nuclei $^{76}G_e$
can meet the experimental data as long as the lightest $\tau$
sneutrino has a mass less than $1{\rm TeV}$. As the lightest
$\tau$ sneutrino mass increases to a certain value, the
theoretically calculated values of the $(\beta\beta)_{0\nu}$ decay
half life time of nuclei $^{76}G_e$ are larger than the
experimental upper limit. When we set the lepton CKM matrix as
${\bf U}_M={\bf U}_2$,  and the other parameters  to be the same
as in Fig.\ref{fig4}, we re-calculate the relation of the
half-life time versus $m_{_{\tilde{\nu}}}$ and the results are
shown in Fig.\ref{fig5}. The trend observed in Fig.\ref{fig5} is
similar to that in Fig.\ref{fig4} and the similarity is understood
as a property of the model.

\section{Discussion and Conclusion}

In this work, we analyze the neutrinoless double beta decay of
nuclei $^{76}G_e$ in the supersymmetric seesaw model. With some
assumptions on the model parameter space, we can naturally explain
the solar and atmospheric neutrino experiments with the
oscillation scenario and simultaneously the HEIDELBERG-MOSCOW
experimental result of the $^{76}G_e$ neutrinoless double decay.
In the numerical analysis, we assume that the masses of the three
generation right handed neutrinos are degenerate, i.e.
$m_{_1}^n=m_{_2}^n=m_{_3}^n =m_{_R}$, the assumption can simplify
our calculation to a certain extent. Meanwhile, we take
$h_1^\nu=0,\;h_2^\nu={h_3^\nu\over 10}$. The choice corresponds to
the hierarchical class solutions of neutrino masses in the
three-generation models. As for the lepton CKM matrix, we choose
two typical mixing matrices. In fact, we may have some other choices on
the supersymmetric seesaw model parameter space to understand the
neutrino oscillation together with the HEIDELBERG-MOSCOW
neutrinoless double beta decay experiment. An important point is
noted that for each special assumption, we must have at least one
sneutrino of middle energy scale, otherwise the data of the three
experiments cannot be simultaneously fitted in this supersymmetric
sea saw model.

Being more explicit, in the preferred three-generation neutrino
model, the contribution from the virtual Majorana-type neutrinos
is suppressed by a factor $<m_{\nu}>/p_F$ of about $10^{-10}$ and
as well as the typical  loop suppression factor at the
electro-weak scale. Thus the SM with right-handed neutrinos cannot
meet the data because of the unavoidable suppression factor. In
contrast, as discussed in the text, the lightest sneutrino has a
Majorana-type component which violates the lepton number and
contributes to the neutrinoless double beta decay and it does not
suffer from the strong suppression factor $<m_{\nu}>/p_F$.
Generally, the fraction of this component is about $10^{-6}$. In
addition to the typical loop suppression, the total suppression
factor is about $10^{-8}$ at the amplitude level. This value (or
just the order of magnitude) can meet the data of the solar and
atmospheric neutrino experiments and the HEIDELBERG-MOSCOW
$(\beta\beta)_{0\nu}$ data simultaneously. Therefore the result
implies that the observed data in the solar and atmospheric
neutrino experiments can be explained by the oscillations
mechanisms of the three-generation neutrinos,  whereas the
neutrinoless double beta decay  is due to the contribution from
the SUSY particles and the applied model is the SUSY see-saw model
with the right-handed neutrinos. On other side, thus one can
expect that the $(\beta\beta)_{0\nu}$ decay data greatly constrain
the parameter space of the model, even though there still is large
free room in the space which should be restricted by further
experiments.

Our conclusion is that the chosen parameter space of the
supersymmetric model with right-handed neutrinos can naturally
explain the observed solar, atmospheric neutrino experiments and
the HEIDELBERG-MOSCOW $(\beta\beta)_{0\nu}$ data. Even though the
parameter space of the model still cannot be completely
determined, the progress is noticeable.

Introducing a "sterile" neutrino or a CPT-violation term in the
neutrino sector, we can also accommodate all the lepton flavor
violation processes\cite{261}. How to distinguish between the
supersymmetric seesaw model and those models is an interesting
subject. Because the "sterile" neutrino participates in the weak
interaction only via the mixing with the normal neutrinos,
accurate measurements and systematic analysis on the $\tau,\;\mu$
rare decays may indicate the difference between the supersymmetry
seesaw and "sterile" neutrino models. One plausible measurement
which can distinguish the supersymmetry seesaw model from the
CPT-violation model is the detection of neutrino magnetic moments
in accurate experimental measurements (if possible). Because in
the SUSY seesaw model, the neutrinos are purely Majorana-type
which cannot have non-zero magnetic moments\cite{042}, by
contraries, the neutrinos which reside in the CPT violation model
can possess a Dirac-type neutrino component, thus can accommodate
a non-zero magnetic moment. Obviously, such measurements would be
extremely difficult, however, on the other side, almost all
experiments on neutrinos are very difficult, so we lay our hope on
the future developments of physics and technology.

\vspace{1.0cm} \noindent {\Large{\bf Acknowledgments}}

This work is partially supported by the National Natural Science
Foundation of China. One of the authors (T.-F. Feng) is also
partly supported by the China postdoctoral foundation and K. C.
Wont Post-doctoral Research fund, Hongkong. We thank Prof. T. Hang
and X.-M. Zany for the helpful discussion on this topic.

\vspace{2cm}
\vspace{0.5cm}\appendix

\section{The sneutrino mixing under MIFF assumption
\label{appa}}

With the MIFF assumption, the sneutrino mass matrix reduces to
(one-generation)
\begin{eqnarray}
{\bf m}_{\tilde{\nu}^I}=\left(\begin{array}{ccc}
\xi_{_I}&\rho_{_I}&\rho_{_I}\\
\rho_{_I}&{1\over 2}m_{_{n^I}}^2+h_{_I}^{\nu^2}\upsilon_2^2
+{1\over 2}m_{_{N^{I}}}^2&{1\over 2} B^n_{_I}\\
\rho_{_I}&{1\over 2} B^n_{_I}&{1\over 2}m_{_{n^I}}^2
+h_{_I}^{\nu^2}\upsilon_2^2 +{1\over
2}m_{_{N^{I}}}^2\end{array}\right)
\end{eqnarray}
with
\begin{eqnarray}
&&\xi_{_I}=-{1\over 2}m_{_{\rm W}}^2\frac{C_{_\beta}^2-S_{_{_\beta}}^2}
{C_{_{\rm W}}^2}+m_{_{\tilde{L}^I}}^2\;,\nonumber\\
&&\rho_{_I}=-\frac{h_{_I}^\nu\mu\upsilon_1+h_{_I}^\nu m_{_{n^I}}\upsilon_2}
{\sqrt{2}}+\frac{A_{_I}^n\upsilon_2}{\sqrt{2}}.
\end{eqnarray}
Defining a symbol
$$\Delta=({1\over 2}m_{_{n^I}}^2+h_{_I}^{\nu^2}\upsilon_2^2
+{1\over 2}m_{_{N^{I}}}^2+{1\over 2} B^n_{_I}-\xi_{_I})^2
+8\rho_{_I}^2,$$ the three eigenvalues of sneutrino mass matrix
are obtained as
\begin{eqnarray}
&&m_{_{\tilde{\nu}_1^I}}^2={1\over 4}\bigg(
m_{_{n^I}}^2+2h_{_I}^{\nu^2}\upsilon_2^2 +m_{_{N^{I}}}^2+
B^n_{_I}+2\xi_{_I}-
2\Delta^{1\over 2}\bigg)\;,\nonumber\\
&&m_{_{\tilde{\nu}_2^I}}^2={1\over 4}\bigg(
m_{_{n^I}}^2+2h_{_I}^{\nu^2}\upsilon_2^2 +m_{_{N^{I}}}^2+
B^n_{_I}+2\xi_{_I}+
2\Delta^{1\over 2}\bigg)\;,\nonumber\\
&&m_{_{\tilde{\nu}_3^I}}^2={1\over 4}\bigg(
m_{_{n^I}}^2+2h_{_I}^{\nu^2}\upsilon_2^2 +m_{_{N^{I}}}^2-
B^n_{_I}\bigg)\;. \label{eigenvalue}
\end{eqnarray}
Assuming the relations $\xi_{_I},\;\rho_{_I}\ll
m_{_{n^I}}^2+2h_{_I}^{\nu^2}\upsilon_2^2 +m_{_{N^{IJ}}}^2$ or
$\xi_{_I},\;\rho_{_I}\ll B^n_{_I}$, Eq.\ref{eigenvalue} is
simplified as (up to order ${\cal
O}(\frac{\rho_{_I},\xi_{_I}}{m_{_{n^I}}^2+2h_{_I}^{\nu^2}
\upsilon_2^2+m_{_{N^{I}}}^2\mp B^n_{_I}})$)
\begin{eqnarray}
&&m_{_{\tilde{\nu}_1^I}}^2=\xi_{_I}-\frac{2\rho_{_I}^2}
{m_{_{\tilde{\nu}_2^I}}^2}\;,\nonumber\\
&&m_{_{\tilde{\nu}_2^I}}^2={1\over 2}\bigg(
m_{_{n^I}}^2+2h_{_I}^{\nu^2}\upsilon_2^2
+m_{_{N^{I}}}^2+ B^n_{_I}\bigg)\;,\nonumber\\
&&m_{_{\tilde{\nu}_3^I}}^2={1\over 4}\bigg(
m_{_{n^I}}^2+2h_{_I}^{\nu^2}\upsilon_2^2 +m_{_{N^{I}}}^2-
B^n_{_I}\bigg)\;. \label{simpeig}
\end{eqnarray}
The corresponding unitary matrix which  dragonflies the mass
matrix is also simplified as Eq.\ref{snumix}.

\section{The expression for $\eta_i$\label{appb}}
\begin{eqnarray}
&&\eta_{a_1}(\mu_{_{\rm W}})=\frac{m_{_{\kappa_l^0}}m_P}
{m_{_{\rm W}}^2}\frac{\alpha_e}{4\pi
S_{_{\rm W}}^2C_{_{\rm W}}^2}\Big(V_{ud}^*\Big)^2U_{mn}U_{op}
({\cal Z}_{_\nu}^{(3+m)j}+{\cal Z}_{_\nu}^{(6+m)j})
({\cal Z}_{_\nu}^{(3+o)j}+{\cal Z}_{_\nu}^{(6+o)j})
{\cal Z}_{_{\tilde L}}^{nk}
{\cal Z}_{_{\tilde L}}^{pi}{\cal Z}_{_{\tilde L}}^{1i*}
{\cal Z}_{_{\tilde L}}^{1k*}
\nonumber \\
&&\hspace{1.8cm}
\Big(S_{_{\rm W}}{\cal Z}_{_{\cal N}}^{1l*}+C_{_{\rm W}}
{\cal Z}_{_{\cal N}}^{2l*}\Big)^2
\sum\limits_{a={\tilde L}_i,{\tilde \nu_j},{\tilde L}_k,\kappa_l^0
}\frac{x_a^2\ln x_a}{\prod\limits_{a\neq b}
(x_b-x_a)}\;,\nonumber \\
&&\eta_{a_2}(\mu_{_{\rm W}})=\frac{m_{_{\kappa_l^0}}m_P}
{m_{_{\rm W}}^2}\frac{\alpha_e}{\pi
C_{_{\rm W}}^2}\Big(V_{ud}^*\Big)^2U_{mn}U_{op}
({\cal Z}_{_\nu}^{(3+m)j}+{\cal Z}_{_\nu}^{(6+m)j})
({\cal Z}_{_\nu}^{(3+o)j}+{\cal Z}_{_\nu}^{(6+o)j})
{\cal Z}_{_{\tilde L}}^{nk}
{\cal Z}_{_{\tilde L}}^{pi}{\cal Z}_{_{\tilde L}}^{4i*}
{\cal Z}_{_{\tilde L}}^{4k*}\nonumber\\
&&\hspace{1.8cm}\Big({\cal Z}_{_{\cal N}}^{1l}\Big)^2
\sum\limits_{a={
\tilde L}_i,{\tilde \nu_j},{\tilde L}_k,\kappa_l^0
}\frac{x_a^2\ln x_a}{\prod\limits_{a\neq b}
(x_b-x_a)}\;,\nonumber \\
&&\eta_{b_1}(\mu_{_{\rm W}})=\frac{2m_P}{m_{_{\rm W}}^4}\frac{\alpha_e}
{4\pi S_{_{\rm W}}^2}U_{em}U_{en}
({\cal Z}_{_\nu}^{(3+m)j*}+{\cal Z}_{_\nu}^{(6+m)j*})
({\cal Z}_{_\nu}^{(3+n)j*}+{\cal Z}_{_\nu}^{(6+n)j*})
\Big({\cal Z}_+^{1i}-\frac{h_m^\nu}{g_2}{\cal Z}_+^{2i}\Big)
\nonumber \\
&&\hspace{1.8cm}\Big({\cal Z}_+^{1k}-\frac{h_n^\nu}{g_2}
{\cal Z}_+^{2k}\Big)\sum\limits_{a=\kappa_i^-,\kappa_j^0,
\kappa_k^-,\tilde{\nu}_{_l}}\frac{x_a\ln x_a}{\prod\limits_{b
\neq a}(x_b-x_a)}\bigg(
m_a^2 m_{_{\kappa_i^-}}\Big({\cal Z}_{_{\cal N}}^{2j}
{\cal Z}_+^{1i*}-\frac{1}{\sqrt{2}}{\cal Z}_{_{\cal N}}^{4j}
{\cal Z}_+^{2i*}\Big)\nonumber \\
&&\hspace{1.8cm}\Big({\cal Z}_{_{\cal N}}^{2j}
{\cal Z}_+^{1k*}-\frac{1}{\sqrt{2}}{\cal Z}_{_{\cal N}}^{4j}
{\cal Z}_+^{2k*}\Big)+m_a^2 m_{_{\kappa_k^-}}\Big({\cal Z}_{_{\cal N}}^{2j*}
{\cal Z}_-^{1i}+\frac{1}{\sqrt{2}}{\cal Z}_{_{\cal N}}^{3j*}
{\cal Z}_+^{2i}\Big)\Big({\cal Z}_{_{\cal N}}^{2j*}
{\cal Z}_-^{1k}\nonumber \\
&&\hspace{1.8cm}+\frac{1}{\sqrt{2}}{\cal Z}_{_{\cal N}}^{3j*}
{\cal Z}_-^{2k}\Big)+4m_{_{\kappa_i^-}}m_{_{\kappa_j^0}}
m_{_{\kappa_k^-}}\Big({\cal Z}_{_{\cal N}}^{2j}
{\cal Z}_+^{1i*}-\frac{1}{\sqrt{2}}{\cal Z}_{_{\cal N}}^{4j}
{\cal Z}_+^{2i*}\Big)\Big({\cal Z}_{_{\cal N}}^{2j*}
{\cal Z}_-^{1k}+\frac{1}{\sqrt{2}}{\cal Z}_{_{\cal N}}^{3j*}
{\cal Z}_-^{2k}\Big)\bigg)\;,
\nonumber \\
&&\eta_{b_2}(\mu_{_{\rm W}})=-\frac{4m_P}{m_{_{\rm W}}^4}\frac{\alpha_e}
{4\pi S_{_{\rm W}}^2}U_{em}U_{en}
({\cal Z}_{_\nu}^{(3+m)j*}+{\cal Z}_{_\nu}^{(6+m)j*})
({\cal Z}_{_\nu}^{(3+n)j*}+{\cal Z}_{_\nu}^{(6+n)j*})
\Big({\cal Z}_+^{1i}-\frac{h_m^\nu}{g_2}{\cal Z}_+^{2i}\Big)
\nonumber \\
&&\hspace{1.8cm}\Big(
{\cal Z}_+^{1k}-\frac{h_n^\nu}{g_2}
{\cal Z}_+^{2k}\Big)\Big({\cal Z}_{_{\cal N}}^{2j*}
{\cal Z}_-^{1i}+\frac{1}{\sqrt{2}}{\cal Z}_{_{\cal N}}^{3j*}
{\cal Z}_+^{2i}\Big)\Big({\cal Z}_{_{\cal N}}^{2j}
{\cal Z}_+^{1k*}-\frac{1}{\sqrt{2}}{\cal Z}_{_{\cal N}}^{4j}
{\cal Z}_+^{2k*}\Big)\nonumber \\
&&\hspace{1.8cm}
\sum\limits_{a=\kappa_i^-,\kappa_j^0,
\kappa_k^-,\tilde{\nu}_{_l}}\frac{x_a\ln x_a}{\prod\limits_{b
\neq a}(x_b-x_a)}m_a^2m_{_{\kappa_j^0}}\;,
\nonumber \\
&&\eta_{c_1}(\mu_{_{\rm W}})=\frac{m_P}{m_{_{\rm W}}^2}\frac{\alpha_e}{4\pi
S_{_{\rm W}}^2}\Big(V_{ud}^*\Big)^2
U_{em}U_{no}({\cal Z}_{_\nu}^{(3+m)l*}+{\cal Z}_{_\nu}^{(6+m)l*})
({\cal Z}_{_\nu}^{(3+n)l}+{\cal Z}_{_\nu}^{(6+n)l})
{\cal Z}_{_{\tilde L}}^{ok}\Big({\cal Z}_+^{1i}-\frac{h_m^\nu}{g_2}
{\cal Z}_+^{2i}\Big)\nonumber\\&&\hspace{1.8cm}
\sum\limits_{a=\kappa_i^-,\kappa_j^0,\tilde{L}_k,\tilde{\nu}_l}
\frac{x_a^2\ln x_a}{\prod\limits_{b\neq a}(x_b-x_a)}
\frac{m_{_{\kappa_j^0}}}{C_{_{\rm W}}}{\cal Z}_{_{\tilde L}
}^{1k*}\Big({\cal Z}_{_{\cal N}}^{2j*}
{\cal Z}_-^{1i}+\frac{1}{\sqrt{2}}{\cal Z}_{_{\cal N}}^{3j*}
{\cal Z}_-^{2i}\Big)\Big(S_{_{\rm W}}{\cal Z}_{_{\cal N}}^{1j*}
+C_{_{\rm W}}{\cal Z}_{_{\cal N}}^{2j*}\Big)\;,\nonumber \\
&&\eta_{c_2}(\mu_{_{\rm W}})=\frac{m_P}{m_{_{\rm W}}^2}\frac{\alpha_e}{4\pi
S_{_{\rm W}}^2}\Big(V_{ud}^*\Big)^2
U_{em}U_{no}({\cal Z}_{_\nu}^{(3+m)l*}+{\cal Z}_{_\nu}^{(6+m)l*})
({\cal Z}_{_\nu}^{(3+n)l}+{\cal Z}_{_\nu}^{(6+n)l})
{\cal Z}_{_{\tilde L}}^{ok}\Big({\cal Z}_+^{1i}-\frac{h_m^\nu}{g_2}
{\cal Z}_+^{2i}\Big)\nonumber\\&&\hspace{1.8cm}
\sum\limits_{a=\kappa_i^-,\kappa_j^0,\tilde{L}_k,\tilde{\nu}_l}
\frac{x_a^2\ln x_a}{\prod\limits_{b\neq a}(x_b-x_a)}
\frac{m_{_{\kappa_i^-}}}{C_{_{\rm W}}}{\cal Z}_{_{\tilde L}
}^{1k*}\Big({\cal Z}_{_{\cal N}}^{2j}
{\cal Z}_-^{2i*}-\frac{1}{\sqrt{2}}{\cal Z}_{_{\cal N}}^{4j}
{\cal Z}_-^{2i*}\Big)\Big(S_{_{\rm W}}{\cal Z}_{_{\cal N}}^{1j*}
+C_{_{\rm W}}{\cal Z}_{_{\cal N}}^{2j*}\Big)\;,\nonumber \\
&&\eta_{d_1}(\mu_{_{\rm W}})=-\frac{m_Pm_{_{\kappa^0_m}}}{m_{_{\rm W}}^2}
\frac{\alpha_e}{8\pi S_{_{\rm W}}^2C_{_{\rm W}}}
(V_{ud}^*)^2U_{en}U_{eo}{\cal Z}_{_{\tilde U}}^{1i}
{\cal Z}_{_{\tilde U}}^{1i*}
({\cal Z}_{_\nu}^{(3+n)k}+{\cal Z}_{_\nu}^{(6+n)k})
({\cal Z}_{_\nu}^{(3+o)k}+{\cal Z}_{_\nu}^{(6+o)k})
{\cal Z}_+^{1j}\nonumber\\&&\hspace{1.8cm}
\Big({\cal Z}_+^{1l*}-\frac{h_n^\nu}{g_2}
{\cal Z}_+^{2l*}\Big)\Big({\cal Z}_+^{1j*}-\frac{h_n^\nu}
{g_2}{\cal Z}_+^{2j*}\Big)
\Big(\frac{1}{3}S_{_{\rm W}}
{\cal Z}_{_{\cal N}}^{1m*}+C_{_{\rm W}}{\cal Z}_{_{\cal N}}^{2m*}
\Big)\Big({\cal Z}_{_{\cal N}}^{2m*}
{\cal Z}_+^{1l}-\frac{1}{\sqrt{2}}{\cal Z}_{_{\cal N}}^{4m*}
{\cal Z}_+^{2l}\Big)\nonumber\\&&\hspace{1.8cm}
\sum\limits_{a={
\tilde U}_i,\kappa_j^-,{\tilde \nu_k},\kappa_l^-,
\kappa_m^0}\frac{x_a^2\ln x_a}{\prod\limits_{a\neq b}
(x_b-x_a)}\;,
\nonumber \\
&&\eta_{d_2}(\mu_{_{\rm W}})=-\frac{m_Pm_{_{\kappa^-_l}}}
{m_{_{\rm W}}^2}\frac{\alpha_e}{6\pi
S_{_{\rm W}}C_{_{\rm W}}}(V_{ud}^*)^2U_{en}U_{eo}
{\cal Z}_{_{\tilde U}}^{1i*}{\cal Z}_{_{\tilde U}}^{4i}
({\cal Z}_{_\nu}^{(3+n)k}+{\cal Z}_{_\nu}^{(6+n)k})
({\cal Z}_{_\nu}^{(3+o)k}+{\cal Z}_{_\nu}^{(6+o)k}){\cal Z}_+^{1j}
\nonumber\\&&\hspace{1.8cm}
\Big({\cal Z}_+^{1l*}-\frac{h_n^\nu}{g_2}{\cal Z}_+^{2l*}\Big)\Big(
{\cal Z}_+^{1j*}-\frac{h_n^\nu}{g_2}{\cal Z}_+^{2j*}\Big)
{\cal Z}_{_{\cal N}}^{1m}
\Big({\cal Z}_{_{\cal N}}^{2m}{\cal Z}_+^{1l*}
+\frac{1}{\sqrt{2}}{\cal Z}_{_{\cal N}}^{3m}
{\cal Z}_+^{2l*}\Big)\sum\limits_{a={
\tilde U}_i,\kappa_j^-,{\tilde \nu_k},\kappa_l^-,
\kappa_m^0}\frac{x_a^2\ln x_a}{\prod\limits_{a\neq b}
(x_b-x_a)}\;,
\nonumber \\
&&\eta_{e_1}(\mu_{_{\rm W}})=\frac{m_Pm_{_{\kappa_m^0}}}{m_{_{\rm W}}^2}
\frac{\alpha_e}{3\pi C_{_{\rm W}}^2}(V_{ud}^*)^2U_{po}U_{en}
{\cal Z}_{_{\tilde U}}^{1i*}{\cal Z}_+^{1j}{\cal Z}_{_{\tilde L}}
^{pl*}{\cal Z}_{_{\tilde U}}
^{4i}\Big({\cal Z}_{_{\cal N}}^{1m}\Big)^2{\cal Z}_{_{\tilde L}}^{4l*}
({\cal Z}_{_\nu}^{(3+n)k}+{\cal Z}_{_\nu}^{(6+n)k})
({\cal Z}_{_\nu}^{(3+o)k}+{\cal Z}_{_\nu}^{(6+o)k})
\nonumber \\
&&\hspace{1.8cm}
\Big({\cal Z}_+^{1j*}-\frac{h_n^\nu}{g_2}{\cal Z}_+^{2j*}\Big)
\sum\limits_{a={
\tilde U}_i,\kappa_j^-,{\tilde \nu_k},\kappa_l^-,
\kappa_m^0}\frac{x_a^2\ln x_a}{\prod\limits_{a\neq b}
(x_b-x_a)}\;,\nonumber \\
&&\eta_{e_2}(\mu_{_{\rm W}})=\frac{m_Pm_{_{\kappa_m^0}}}{m_{_{\rm W}}^2}
\frac{\alpha_e}{8\pi S_{_{\rm W}}^2C_{_{\rm W}}^2}(V_{ud}^*)^2U_{po}U_{en}
{\cal Z}_{_{\tilde U}}^{1i*}{\cal Z}_+^{1j}{\cal Z}_{_{\tilde L}}
^{pl*}{\cal Z}_{_{\tilde U}}^{1i}{\cal Z}_{_{\tilde L}}^{1l*}
({\cal Z}_{_\nu}^{(3+n)k}+{\cal Z}_{_\nu}^{(6+n)k})
({\cal Z}_{_\nu}^{(3+o)k}+{\cal Z}_{_\nu}^{(6+o)k})
\nonumber\\&&\hspace{1.8cm}
\Big(\frac{1}{3}S_{_{\rm W}}{\cal Z}_{_{\cal N}}^{1m*}
+C_{_{\rm W}}{\cal Z}_{_{\cal N}}^{2m*}\Big)\Big(S_{_{\rm W}}
{\cal Z}_{_{\cal N}}^{1m*}+C_{_{\rm W}}{\cal Z}_{_{\cal N}}^{2m*}
\Big)\Big({\cal Z}_+^{1j*}-\frac{h_n^\nu}{g_2}{\cal Z}_+^{2j*}
\Big)\sum\limits_{a={
\tilde U}_i,\kappa_j^-,{\tilde \nu_k},\kappa_l^-,
\kappa_m^0}\frac{x_a^2\ln x_a}{\prod\limits_{a\neq b}
(x_b-x_a)}\;,\nonumber \\
&&\eta_{f_1}(\mu_{_{\rm W}})=\frac{m_Pm_{_{\kappa_n^0}}}{m_{_{\rm W}}^2}
\frac{4\alpha_e}{9\pi C_{_{\rm W}}^2}(V_{ud}^*)^2U_{eo}U_{ep}{\cal Z}
_{_{\tilde U}}^{1i*}{\cal Z}_{_{\tilde U}}^{1m*}{\cal Z}_+^{1j}
{\cal Z}_+^{1l}{\cal Z}_{_{\tilde U}}^{4i}{\cal Z}_{_{\tilde U}}^{4m}
\Big({\cal Z}_{_{\cal N}}^{1n}\Big)^2
({\cal Z}_{_\nu}^{(3+o)k}+{\cal Z}_{_\nu}^{(6+o)k})
({\cal Z}_{_\nu}^{(3+p)k}\nonumber\\&&\hspace{1.8cm}
+{\cal Z}_{_\nu}^{(6+p)k})
\Big({\cal Z}_+^{1j*}-\frac{h_o^\nu}{g_2}{\cal Z}_+^{2j*}
\Big)\Big({\cal Z}_+^{1l*}-\frac{h_p^\nu}{g_2}
{\cal Z}_+^{2l*}\Big)\sum\limits_{a={
\tilde U}_i,\kappa_j^-,{\tilde \nu_k},\kappa_l^-,{\tilde U_m},
\kappa_n^0}\frac{x_a^2\ln x_a}{\prod\limits_{a\neq b}
(x_b-x_a)}\;,
\nonumber \\
&&\eta_{f_2}(\mu_{_{\rm W}})=\frac{m_Pm_{_{\kappa_n^0}}}{m_{_{\rm W}}^2}
\frac{\alpha_e}{8\pi S_{_{\rm W}}^2}(V_{ud}^*)^2U_{eo}U_{ep}{\cal Z}
_{_{\tilde U}}^{1i*}{\cal Z}_{_{\tilde U}}^{1m*}{\cal Z}_+^{1j}
{\cal Z}_+^{1l}{\cal Z}_{_{\tilde U}}^{1i}
{\cal Z}_{_{\tilde U}}^{1m}
({\cal Z}_{_\nu}^{(3+o)k}+{\cal Z}_{_\nu}^{(6+o)k})
({\cal Z}_{_\nu}^{(3+p)k}+{\cal Z}_{_\nu}^{(6+p)k})
\nonumber\\&&\hspace{1.8cm}
\Big(\frac{1}{3}{\cal Z}_{_{\cal N}}^{1n*}
S_{_{\rm W}}+{\cal Z}_{_{\cal N}}^{2n*}C_{_{\rm W}}\Big)^2
\Big({\cal Z}_+^{1j*}-\frac{h_o^\nu}{g_2}{\cal Z}_+^{2j*}\Big)
\Big({\cal Z}_+^{1l*}-\frac{h_p^\nu}{g_2}{\cal Z}_+^{2l*}\Big)
\sum\limits_{a={
\tilde U}_i,\kappa_j^-,{\tilde \nu_k},\kappa_l^-,{\tilde U_m},
\kappa_n^0}\frac{x_a^2\ln x_a}{\prod\limits_{a\neq b}
(x_b-x_a)}\;,
\nonumber \\
&&\eta_{g_1}(\mu_{_{\rm W}})=\frac{8m_Pm_{_{\tilde g}}}{m_{_{\rm W}}^2}
\frac{\alpha_s}{9\pi}(V_{ud}^*)^2U_{en}U_{eo}{\cal Z}_{_{\tilde U}}
^{1i}{\cal Z}_{_{\tilde U}}^{1m}{\cal Z}_{_{\tilde U}}
^{1i*}{\cal Z}_{_{\tilde U}}^{1m*}{\cal Z}^{1j}_+{\cal Z}^{1l}_+
({\cal Z}_{_\nu}^{(3+n)k}+{\cal Z}_{_\nu}^{(6+n)k})
({\cal Z}_{_\nu}^{(3+o)k}+{\cal Z}_{_\nu}^{(6+o)k})
\nonumber\\&&\hspace{1.8cm}
\Big({\cal Z}^{1j*}_{+}-\frac{h_{n}^{\nu}}{g_2}
{\cal Z}^{2j*}_{+}\Big)\Big({\cal Z}^{1l*}_{+}-\frac{h_{o}^{\nu}}{g_2}
{\cal Z}^{2l*}_{+}\Big)
\sum\limits_{a={\tilde{g},\tilde U}_i,\kappa_j^-,{\tilde \nu_k},
\kappa_l^-,{\tilde U_m}}\frac{x_a^2\ln x_a}{\prod\limits_{a\neq b}
(x_b-x_a)},\nonumber \\
&&\eta_{g_2}(\mu_{_{\rm W}})=\frac{8m_Pm_{_{\tilde g}}}{m_{_{\rm W}}^2}
\frac{\alpha_s}{9\pi}(V_{ud}^*)^2U_{en}U_{eo}{\cal Z}_{_{\tilde U}}
^{4i}{\cal Z}_{_{\tilde U}}^{4m}{\cal Z}_{_{\tilde U}}
^{1i*}{\cal Z}_{_{\tilde U}}^{1m*}{\cal Z}^{1j}_+{\cal Z}^{1l}_+
({\cal Z}_{_\nu}^{(3+n)k}+{\cal Z}_{_\nu}^{(6+n)k})
({\cal Z}_{_\nu}^{(3+o)k}+{\cal Z}_{_\nu}^{(6+o)k})
\nonumber\\&&\hspace{1.8cm}
\Big({\cal Z}^{1j*}_{+}-\frac{h_{n}^{\nu}}{g_2}
{\cal Z}^{2j*}_{+}\Big)\Big({\cal Z}^{1l*}_{+}-\frac{h_{o}^{\nu}}{g_2}
{\cal Z}^{2l*}_{+}\Big)
\sum\limits_{a={\tilde{g},\tilde U}_i,\kappa_j^-,{\tilde \nu_k},
\kappa_l^-,{\tilde U_m}}\frac{x_a^2\ln x_a}{\prod\limits_{a\neq b}
(x_b-x_a)}\;,
\nonumber \\
&&\eta_{h_1}(\mu_{_{\rm W}})=\frac{m_Pm_{_{\kappa_k^-}}}{m_{_{\rm W}}^2}
\frac{\alpha_e}{6\pi S_{_{\rm W}}C_{_{\rm W}}}
\Big(V_{ud}^*\Big)^2U_{en}U_{eo}{\cal Z}_{_{\tilde D}}^{1i*}
{\cal Z}_{_{\tilde D}}^{4i}{\cal Z}_-^{1m*}{\cal Z}_{_{\cal
N}}^{1j*}({\cal Z}_{_\nu}^{(3+n)l*}+{\cal Z}_{_\nu}^{(6+n)l*})
({\cal Z}_{_\nu}^{(3+o)l*}+{\cal Z}_{_\nu}^{(6+o)l*})
\nonumber\\&&\hspace{1.8cm}
\Big({\cal Z}_+^{1m}-\frac{h_n^\nu}{g_2}{\cal Z}_+^{2m}
\Big)\Big({\cal Z}_+^{1k}-\frac{h_o^\nu}{g_2}{\cal Z}_+^{2k}
\Big)\Big({\cal Z}_{_{\cal N}}^{2j}{\cal Z}_+^{1k*}-{1\over \sqrt{2}}
{\cal Z}_{_{\cal N}}^{4j}{\cal Z}_+^{2k*}\Big)
\sum\limits_{a={\tilde{D}_i,\kappa_j^0,\kappa_k^-,{\tilde \nu}_l,
,\kappa_m^-}}\frac{x_a^2\ln x_a}{\prod\limits_{a\neq b}
(x_b-x_a)}\;,\nonumber\\
&&\eta_{h_2}(\mu_{_{\rm W}})=\frac{m_Pm_{_{\kappa_j^0}}}{m_{_{\rm W}}^2}
\frac{\alpha_e}{6\pi S_{_{\rm W}}C_{_{\rm W}}}
\Big(V_{ud}^*\Big)^2U_{en}U_{eo}{\cal Z}_{_{\tilde D}}^{1i*}
{\cal Z}_{_{\tilde D}}^{4i}{\cal Z}_-^{1m*}{\cal Z}_{_{\cal
N}}^{1j*}({\cal Z}_{_\nu}^{(3+n)l*}+{\cal Z}_{_\nu}^{(6+n)l*})
({\cal Z}_{_\nu}^{(3+o)l*}+{\cal Z}_{_\nu}^{(6+o)l*})
\nonumber\\&&\hspace{1.8cm}\Big({\cal Z}_+^{1m}
-\frac{h_n^\nu}{g_2}{\cal Z}_+^{2m}\Big)\Big({\cal Z}_+^{1k}
-\frac{h_o^\nu}{g_2}{\cal Z}_+^{2k}\Big)\Big({\cal Z}_{_{\cal N}}^{2j*}
{\cal Z}_-^{1k}+{1\over \sqrt{2}}
{\cal Z}_{_{\cal N}}^{4j*}{\cal Z}_-^{2k}\Big)
\sum\limits_{a={\tilde{D}_i,\kappa_j^0,\kappa_k^-,{\tilde \nu}_l,
,\kappa_m^-}}\frac{x_a^2\ln x_a}{\prod\limits_{a\neq b}
(x_b-x_a)}\;,\nonumber\\
&&\eta_{i_1}(\mu_{_{\rm W}})=\frac{m_Pm_{_{\kappa_j^0}}}{m_{_{\rm W}}^2}
\frac{\alpha_e}{12\pi S_{_{\rm W}}C_{_{\rm W}}^2}
\Big(V_{ud}^*\Big)^2U_{en}U_{IJ}{\cal Z}_{_{\tilde D}}^{1i*}{\cal Z}
_{_{\tilde D}}^{4i}
{\cal Z}_-^{1m*}{\cal Z}_{_{\cal N}}^{1j*}
{\cal Z}_{_{\tilde L}}^{1l}{\cal Z}_{_{\tilde L}}^{Jl}
({\cal Z}_{_\nu}^{(3+I)k}+{\cal Z}_{_\nu}^{(6+I)k})
\nonumber\\&&\hspace{1.8cm}
({\cal Z}_{_\nu}^{(3+n)k*}+{\cal Z}_{_\nu}^{(6+n)k*})
\Big({\cal Z}_+^{1m}-\frac{h_n^\nu}{g_2}{\cal Z}_+^{2m}
\Big)({\cal Z}_{_{\cal N}}^{1j}S_{_{\rm W}}
+{\cal Z}_{_{\cal N}}^{2j}C_{_{\rm W}})
\sum\limits_{a={\tilde{D}_i,\kappa_j^0,\kappa_k^-,
{\tilde \nu}_l,\kappa_m^-}}\frac{x_a^2\ln x_a}{\prod\limits_{a\neq b}
(x_b-x_a)}\;,\nonumber \\
&&\eta_{i_2}(\mu_{_{\rm W}})=-\frac{m_Pm_{_{\kappa_j^0}}}{m_{_{\rm W}}^2}
\frac{\alpha_e}{4\pi S_{_{\rm W}}C_{_{\rm W}}^2}
\Big(V_{ud}^*\Big)^2U_{en}U_{IJ}{\cal Z}_{_{\tilde D}}^{1i}{\cal Z}
_{_{\tilde D}}^{1i*}{\cal Z}_-^{1m*}{\cal Z}_{_{\cal N}}^{1j*}
{\cal Z}_{_{\tilde L}}^{4l}{\cal Z}_{_{\tilde L}}^{Jl}
({\cal Z}_{_\nu}^{(3+I)k}+{\cal Z}_{_\nu}^{(6+I)k})
\nonumber\\&&\hspace{1.8cm}
({\cal Z}_{_\nu}^{(3+n)k*}+{\cal Z}_{_\nu}^{(6+n)k*})
\Big({\cal Z}_+^{1m}-\frac{h_n^\nu}{g_2}{\cal Z}_+^{2m}\Big)
({1\over 3}{\cal Z}_{_{\cal N}}^{1j}S_{_{\rm W}}
-{\cal Z}_{_{\cal N}}^{2j}C_{_{\rm W}})
\sum\limits_{a={\tilde{D}_i,\kappa_j^0,\kappa_k^-,
{\tilde \nu}_l,\kappa_m^-}}\frac{x_a^2\ln x_a}{\prod\limits_{a\neq b}
(x_b-x_a)}\;,\nonumber \\
&&\eta_{j_1}(\mu_{_{\rm W}})=\frac{m_Pm_{_{\kappa_j^0}}}{m_{_{\rm W}}^2}
\frac{\alpha_e}{12\pi S_{_{\rm W}}^2C_{_{\rm W}}}
\Big(V_{ud}^*\Big)^2U_{eo}U_{ep}{\cal Z}_{
_{\tilde D}}^{1i}{\cal Z}_{_{\tilde D}}^{1k}
{\cal Z}_{_{\tilde D}}^{1i*}{\cal Z}_{_{\tilde D}}^{1k*}
{\cal Z}_-^{1n*}{\cal Z}_-^{1l*}
({\cal Z}_{_\nu}^{(3+o)m*}+{\cal Z}_{_\nu}^{(6+o)m*})
\nonumber\\&&\hspace{1.8cm}
({\cal Z}_{_\nu}^{(3+p)m*}+{\cal Z}_{_\nu}^{(6+p)m*})
\Big({\cal Z}_+^{1n}-\frac{h_o^\nu}{g_2}{\cal Z}_+^{2n}
\Big)\Big({\cal Z}_+^{1l}-\frac{h_p^\nu}{g_2}{\cal Z}_+^{2l}\Big)
\Big({1\over 3}{\cal Z}_{_{\cal N}}^{1j}S_{_{\rm W}}-{\cal Z}_
{_{\cal N}}^{2j}C_{_{\rm W}}\Big)^2\nonumber\\&&\hspace{1.2cm}
\sum\limits_{a={\tilde{D}_i,\kappa_j^0,\tilde{D}_k,\kappa_l^-,
{\tilde \nu}_m,\kappa_n^-}}\frac{x_a^2\ln x_a}{\prod\limits_{a\neq b}
(x_b-x_a)}\;,\nonumber\\
&&\eta_{j_2}(\mu_{_{\rm W}})=\frac{m_Pm_{_{\kappa_j^0}}}{m_{_{\rm W}}^2}
\frac{\alpha_e}{27\pi C_{_{\rm W}}^2}
\Big(V_{ud}^*\Big)^2U_{eo}U_{ep}{\cal Z}_{
_{\tilde D}}^{4i}{\cal Z}_{_{\tilde D}}^{4k}
{\cal Z}_{_{\tilde D}}^{1i*}{\cal Z}_{_{\tilde D}}^{1k*}
{\cal Z}_-^{1n*}{\cal Z}_-^{1l*}
({\cal Z}_{_\nu}^{(3+o)m*}+{\cal Z}_{_\nu}^{(6+o)m*})
\nonumber\\&&\hspace{1.8cm}
({\cal Z}_{_\nu}^{(3+p)m*}+{\cal Z}_{_\nu}^{(6+p)m*})
\Big({\cal Z}_+^{1n}-\frac{h_o^\nu}{g_2}{\cal Z}_+^{2n}
\Big)\Big({\cal Z}_+^{1l}-\frac{h_p^\nu}{g_2}{\cal Z}_+^{2l}
\Big)\Big({\cal Z}_{_{\cal N}}^{1j*}\Big)^2\nonumber \\
&&\hspace{1.8cm}
\sum\limits_{a={\tilde{D}_i,\kappa_j^0,\tilde{D}_k,\kappa_l^-,
{\tilde \nu}_m,\kappa_n^-}}\frac{x_a^2\ln x_a}{\prod\limits_{a\neq b}
(x_b-x_a)}\;,\nonumber\\
&&\eta_{k_1}(\mu_{_{\rm W}})=\frac{m_Pm_{_{\tilde g}}}{m_{_{\rm W}}^2}
\frac{4\alpha_s}{9\pi}\Big(V_{ud}^*\Big)^2U_{eo}U_{ep}
{\cal Z}_{_{\tilde D}}^{1i}{\cal Z}_{_{\tilde D}}^{1k}
{\cal Z}_{_{\tilde D}}^{1i*}{\cal Z}_{_{\tilde D}}^{1k*}
{\cal Z}_-^{1n*}{\cal Z}_-^{1l*}
({\cal Z}_{_\nu}^{(3+o)m*}+{\cal Z}_{_\nu}^{(6+o)m*})
\nonumber\\&&\hspace{1.8cm}
({\cal Z}_{_\nu}^{(3+p)m*}+{\cal Z}_{_\nu}^{(6+p)m*})
\Big({\cal Z}_+^{1n}-\frac{h_o^\nu}{g_2}{\cal Z}_+^{2n}\Big)
\Big({\cal Z}_+^{1l}-\frac{h_p^\nu}{g_2}{\cal Z}_+^{2l}
\Big)\sum\limits_{a={\tilde{D}_i,\tilde{g},
\tilde{D}_k,\kappa_l^-,{\tilde \nu}_m,\kappa_n^-}}\frac{x_a^2\ln x_a}
{\prod\limits_{a\neq b}(x_b-x_a)}\;,\nonumber\\
&&\eta_{k_2}(\mu_{_{\rm W}})=\frac{m_Pm_{_{\tilde g}}}{m_{_{\rm W}}^2}
\frac{4\alpha_s}{9\pi}\Big(V_{ud}^*\Big)^2U_{eo}U_{ep}
{\cal Z}_{_{\tilde D}}^{4i}{\cal Z}_{_{\tilde D}}^{4k}
{\cal Z}_{_{\tilde D}}^{1i*}{\cal Z}_{_{\tilde D}}^{1k*}
{\cal Z}_-^{1n*}{\cal Z}_-^{1l*}
({\cal Z}_{_\nu}^{(3+o)m*}+{\cal Z}_{_\nu}^{(6+o)m*})
\nonumber\\&&\hspace{1.8cm}
({\cal Z}_{_\nu}^{(3+p)m*}+{\cal Z}_{_\nu}^{(6+p)m*})
\Big({\cal Z}_+^{1n}-\frac{h_o^\nu}{g_2}{\cal Z}_+^{2n}\Big)
\Big({\cal Z}_+^{1l}-\frac{h_p^\nu}{g_2}{\cal Z}_+^{2l}
\Big)\sum\limits_{a={\tilde{D}_i,\tilde{g},
\tilde{D}_k,\kappa_l^-,{\tilde \nu}_m,\kappa_n^-}}\frac{x_a^2\ln x_a}
{\prod\limits_{a\neq b}(x_b-x_a)}\;,\nonumber\\
&&\eta_{l_1}(\mu_{_{\rm W}})=\frac{m_Pm_{_{\kappa_j^0}}}{m_{_{\rm W}}^2}
\frac{\alpha_e}{8\pi S_{_{\rm W}}^2C_{_{\rm W}}^2}
\Big(V_{ud}^*\Big)^2U_{eo}U_{ep}{\cal Z}_{_{\tilde U}}^{1l}
{\cal Z}_{_{\tilde D}}^{1i}{\cal Z}_{_{\tilde D}}^{1i*}
{\cal Z}_{_{\tilde U}}^{1l*}{\cal Z}_-^{1n*}{\cal Z}_+^{1k}
({\cal Z}_{_\nu}^{(3+o)m*}+{\cal Z}_{_\nu}^{(6+o)m*})
\nonumber\\&&\hspace{1.8cm}
({\cal Z}_{_\nu}^{(3+p)m}+{\cal Z}_{_\nu}^{(6+p)m})
\Big({\cal Z}_+^{1n}-\frac{h_o^\nu}{g_2}{\cal Z}_+^{2n}\Big)
\Big({\cal Z}_+^{1k*}-\frac{h_p^\nu}{g_2}{\cal Z}_+^{2k*}
\Big)\Big({1\over 3}
{\cal Z}_{_{\cal N}}^{1j*}S_{_{\rm W}}+{\cal Z}_{_{\cal N}}^{2j*}
C_{_{\rm W}}\Big)
\nonumber \\
&&\hspace{1.8cm}
\Big({1\over 3}{\cal Z}_{_{\cal N}}^{1j}S_{_{\rm W}}
-{\cal Z}_{_{\cal N}}^{2j}C_{_{\rm W}}\Big)
\sum\limits_{a={\tilde{D}_i,\kappa_j^0,\kappa_k^-,\tilde{U}_l,
{\tilde \nu}_m,\kappa_n^-}}\frac{x_a^2\ln x_a}{\prod\limits_{a\neq b}
(x_b-x_a)}\;,\nonumber\\
&&\eta_{l_2}(\mu_{_{\rm W}})=-\frac{m_Pm_{_{\kappa_j^0}}}{m_{_{\rm W}}^2}
\frac{\alpha_e}{3\pi C_{_{\rm W}}^2}
\Big(V_{ud}^*\Big)^2U_{eo}U_{ep}{\cal Z}_{_{\tilde U}}^{4l}
{\cal Z}_{_{\tilde D}}^{4i}{\cal Z}_{_{\tilde D}}^{1i*}
{\cal Z}_{_{\tilde U}}^{1l*}{\cal Z}_-^{1n*}{\cal Z}_+^{1k}
({\cal Z}_{_\nu}^{(3+o)m*}+{\cal Z}_{_\nu}^{(6+o)m*})
\nonumber\\&&\hspace{1.8cm}
({\cal Z}_{_\nu}^{(3+p)m}+{\cal Z}_{_\nu}^{(6+p)m})
\Big({\cal Z}_+^{1n}-\frac{h_o^\nu}{g_2}{\cal Z}_+^{2n}\Big)
\Big({\cal Z}_+^{1k*}-\frac{h_p^\nu}{g_2}{\cal Z}_+^{2k*}
\Big){\cal Z}_{_{\cal N}}^{1j*}{\cal Z}_{_{\cal N}}^{1j}
\nonumber\\&&\hspace{1.8cm}
\sum\limits_{a={\tilde{D}_i,\kappa_j^0,\kappa_k^-,\tilde{U}_l,
{\tilde \nu}_m,\kappa_n^-}}\frac{x_a^2\ln x_a}{\prod\limits_{a\neq b}
(x_b-x_a)}\;,\nonumber\\
&&\eta_{m_1}(\mu_{_{\rm W}})=\frac{m_Pm_{_{\tilde g}}}{m_{_{\rm W}}^2}
\frac{4\alpha_s}{9\pi}\Big(V_{ud}^*\Big)^2U_{eo}U_{ep}
{\cal Z}_{_{\tilde U}}^{4l}{\cal Z}_{_{\tilde D}}^{1i}
{\cal Z}_{_{\tilde D}}^{1i*}{\cal Z}_{_{\tilde U}}^{1l*}
{\cal Z}_-^{1n*}{\cal Z}_+^{1k}
({\cal Z}_{_\nu}^{(3+o)m*}+{\cal Z}_{_\nu}^{(6+o)m*})
({\cal Z}_{_\nu}^{(3+p)m}+{\cal Z}_{_\nu}^{(6+p)m})
\nonumber\\&&\hspace{1.8cm}
\Big({\cal Z}_+^{1n}-\frac{h_o^\nu}{g_2}{\cal Z}_+^{2n}
\Big)\Big({\cal Z}_+^{1k*}
-\frac{h_p^\nu}{g_2}{\cal Z}_+^{2k*}\Big)
\sum\limits_{a={\tilde{D}_i,\tilde{g},\kappa_k^-,\tilde{U}_l,
{\tilde \nu}_m,\kappa_n^-}}\frac{x_a^2\ln x_a}{\prod\limits_{a\neq b}
(x_b-x_a)}\;,\nonumber\\
&&\eta_{m_2}(\mu_{_{\rm W}})=\frac{m_Pm_{_{\tilde g}}}{m_{_{\rm W}}^2}
\frac{4\alpha_s}{9\pi}\Big(V_{ud}^*\Big)^2U_{eo}U_{ep}
{\cal Z}_{_{\tilde U}}^{1l}{\cal Z}_{_{\tilde D}}^{4i}
{\cal Z}_{_{\tilde D}}^{1i*}{\cal Z}_{_{\tilde U}}^{1l*}
{\cal Z}_-^{1n*}{\cal Z}_+^{1k}
({\cal Z}_{_\nu}^{(3+o)m*}+{\cal Z}_{_\nu}^{(6+o)m*})
({\cal Z}_{_\nu}^{(3+p)m}+{\cal Z}_{_\nu}^{(6+p)m})
\nonumber\\&&\hspace{1.8cm}
\Big({\cal Z}_+^{1n}-\frac{h_o^\nu}{g_2}{\cal Z}_+^{2n}
\Big)\Big({\cal Z}_+^{1k*}-\frac{h_p^\nu}{g_2}{\cal Z}_+^{2k*}\Big)
\sum\limits_{a={\tilde{D}_i,\tilde{g},\kappa_k^-,\tilde{U}_l,
{\tilde \nu}_m,\kappa_n^-}}\frac{x_a^2\ln x_a}{\prod\limits_{a\neq b}
(x_b-x_a)}\;.
\label{eta}
\end{eqnarray}
Where $x_{i}=\frac{m_i^2}{m_{_{\rm W}}^2}$. The coefficient
$C_i(\mu_{_{\rm W}})\;(i=1,\;2,\;\cdots,\;14)$ are defined as
\begin{eqnarray}
&&C_1(\mu_{_{\rm W}})=\eta_{a_1}(\mu_{_{\rm W}})+\eta_{b_1}(\mu_{_{\rm W}})
+\eta_{b_2}(\mu_{_{\rm W}})+\eta_{c_1}(\mu_{_{\rm W}})
+\eta_{c_2}(\mu_{_{\rm W}})+\eta_{e_2}(\mu_{_{\rm W}})
+\eta_{f_2}(\mu_{_{\rm W}})+\eta_{g_1}(\mu_{_{\rm W}})\;,\nonumber\\
&&C_2(\mu_{_{\rm W}})=\eta_{a_2}(\mu_{_{\rm W}})+\eta_{i_2}(\mu_{_{\rm W}})
+\eta_{j_1}(\mu_{_{\rm W}})+\eta_{k_1}(\mu_{_{\rm W}})\;,\nonumber\\
&&C_3(\mu_{_{\rm W}})=2\eta_{h_1}(\mu_{_{\rm W}})-\eta_{h_2}(\mu_{_{\rm W}})
+\eta_{i_1}(\mu_{_{\rm W}})+\eta_{l_2}(\mu_{_{\rm W}})
+\eta_{m_2}(\mu_{_{\rm W}})\;,
\nonumber\\
&&C_4(\mu_{_{\rm W}})=\eta_{e_1}(\mu_{_{\rm W}})+\eta_{l_1}(\mu_{_{\rm W}})
+\eta_{m_1}(\mu_{_{\rm W}})\;,\nonumber\\
&&C_5(\mu_{_{\rm W}})=-{\eta_{d_1}(\mu_{_{\rm W}})\over 2}
+\eta_{d_2}(\mu_{_{\rm W}})\;,\nonumber\\
&&C_6(\mu_{_{\rm W}})=\eta_{f_1}(\mu_{_{\rm W}})
+\eta_{g_2}(\mu_{_{\rm W}})\;,\nonumber\\
&&C_7(\mu_{_{\rm W}})=\eta_{j_2}(\mu_{_{\rm W}})
+\eta_{k_2}(\mu_{_{\rm W}})\;,\nonumber\\
&&C_8(\mu_{_{\rm W}})=\eta_{b_1}(\mu_{_{\rm W}})
+\eta_{c_2}(\mu_{_{\rm W}})-\eta_{c_1}(\mu_{_{\rm W}})+{1\over
4}\eta_{e_2}(\mu_{_{\rm W}})\;,\nonumber\\
&&C_9(\mu_{_{\rm W}})=-\eta_{i_2}(\mu_{_{\rm W}})
+{1\over 2}\eta_{e_1}(\mu_{_{\rm W}})\;,\nonumber\\
&&C_{10}(\mu_{_{\rm W}})={7\over 4}\eta_{d_1}(\mu_{_{\rm W}})
-\eta_{d_2}(\mu_{_{\rm W}})\;,\nonumber\\
&&C_{11}(\mu_{_{\rm W}})=-{1\over 2}\Big(\eta_{e_1}(\mu_{_{\rm W}})
+\eta_{i_2}(\mu_{_{\rm W}})+3\eta_{l_1}(\mu_{_{\rm W}})
+3\eta_{m_1}(\mu_{_{\rm W}})\Big)\;,\nonumber\\
&&C_{12}(\mu_{_{\rm W}})=-\Big(\eta_{h_2}(\mu_{_{\rm W}})
-{3\over 4}\eta_{h_1}(\mu_{_{\rm W}})+{1\over 4}\eta_{i_1}(\mu_{_{\rm W}})
-{1\over 2}\eta_{l_2}(\mu_{_{\rm W}})
-{1\over 2}\eta_{m_2}(\mu_{_{\rm W}})\Big)\;,\nonumber\\
&&C_{13}(\mu_{_{\rm W}})={1\over 4}\Big(\eta_{l_1}(\mu_{_{\rm W}})
+\eta_{m_1}(\mu_{_{\rm W}})+\eta_{l_2}(\mu_{_{\rm W}})
+\eta_{m_2}(\mu_{_{\rm W}})\Big)
\;,\nonumber\\
&&C_{14}(\mu_{_{\rm W}})=-{1\over 4}\Big(\eta_{h_1}(\mu_{_{\rm W}})
+\eta_{i_1}(\mu_{_{\rm W}})+\eta_{i_2}(\mu_{_{\rm W}})\Big)\;.
\label{ci}
\end{eqnarray}


\end{document}